\documentclass[apj,iop]{emulateapj}
\usepackage{color}
\usepackage{times}
\usepackage{hyperref}
\hypersetup{
    pdftitle={The SN Ia Rate in High-Redshift Galaxy Clusters},
    pdfauthor={Kyle Barbary},
    colorlinks=true,        
    linkcolor=blue,         
    citecolor=blue,         
    urlcolor=blue           
}
\slugcomment{Accepted for publication in the Astrophysical Journal on 16 February 2011}
\shorttitle{The SN~Ia Rate in High-Redshift Galaxy Clusters}
\shortauthors{Barbary et al.}
\begin{document}
\title{The {\it Hubble Space Telescope}\footnotemark[*] 
       Cluster Supernova Survey:\\
       II. The Type I\lowercase{a} Supernova Rate in High-Redshift Galaxy Clusters}
\footnotetext[*]{Based in part on observations made with the 
  NASA/ESA {\it Hubble Space Telescope}, obtained from the data
  archive at the Space Telescope Institute. STScI is operated by the
  association of Universities for Research in Astronomy, Inc. under
  the NASA contract NAS 5-26555.  The observations are associated with
  program GO-10496.}
\author{
K.~Barbary\altaffilmark{1,2},
G.~Aldering\altaffilmark{2},
R.~Amanullah\altaffilmark{1,3},
M.~Brodwin\altaffilmark{4,5},
N.~Connolly\altaffilmark{6},
K.~S.~Dawson\altaffilmark{2,7},
M.~Doi\altaffilmark{8},
P.~Eisenhardt\altaffilmark{9},
L.~Faccioli\altaffilmark{2},
V.~Fadeyev\altaffilmark{10},
H.~K.~Fakhouri\altaffilmark{1,2},
A.~S.~Fruchter\altaffilmark{11},
D.~G.~Gilbank \altaffilmark{12},
M.~D.~Gladders\altaffilmark{13},
G.~Goldhaber\altaffilmark{1,2,26}, 
A.~Goobar\altaffilmark{3,14},
T.~Hattori\altaffilmark{15},
E.~Hsiao\altaffilmark{2},
X.~Huang\altaffilmark{1},
Y.~Ihara\altaffilmark{8,25},
N.~Kashikawa\altaffilmark{16},
B.~Koester\altaffilmark{13,17},
K.~Konishi\altaffilmark{18},
M.~Kowalski\altaffilmark{19},
C.~Lidman\altaffilmark{20},
L.~Lubin\altaffilmark{21},
J.~Meyers\altaffilmark{1,2},
T.~Morokuma\altaffilmark{8,16,25},
T.~Oda\altaffilmark{22},
N.~Panagia\altaffilmark{11},
S.~Perlmutter\altaffilmark{1,2},
M.~Postman\altaffilmark{11},
P.~Ripoche\altaffilmark{2},
P.~Rosati\altaffilmark{23},
D.~Rubin\altaffilmark{1,2},
D.~J.~Schlegel\altaffilmark{2},
A.~L.~Spadafora\altaffilmark{2},
S.~A.~Stanford\altaffilmark{21,24},
M.~Strovink\altaffilmark{1,2},
N.~Suzuki\altaffilmark{2},
N.~Takanashi\altaffilmark{16},
K.~Tokita\altaffilmark{8},
N.~Yasuda\altaffilmark{18} \\
(The Supernova Cosmology Project)
}
\email{kbarbary@lbl.gov}
\altaffiltext{1}{Department of Physics, University of California,
Berkeley, CA 94720}
\altaffiltext{2}{E.~O. Lawrence Berkeley National Lab, 1 Cyclotron Rd.,
Berkeley, CA 94720}
\altaffiltext{3}{The Oskar Klein Centre for Cosmo Particle Physics,
AlbaNova, SE-106 91 Stockholm, Sweden}
\altaffiltext{4}{Harvard-Smithsonian Center for Astrophysics, 60
Garden Street, Cambridge, MA 02138}
\altaffiltext{5}{W. M. Keck Postdoctoral Fellow at the
Harvard-Smithsonian Center for Astrophysics}
\altaffiltext{6}{Hamilton College Department of Physics, Clinton, NY 13323}
\altaffiltext{7}{Department of Physics and Astronomy, University of
Utah, Salt Lake City, UT 84112}
\altaffiltext{8}{Institute of Astronomy, Graduate School of Science,
University of Tokyo 2-21-1 Osawa, Mitaka, Tokyo 181-0015, Japan}
\altaffiltext{9}{Jet Propulsion Laboratory, California Institute of
Technology, Pasadena, CA 91109}
\altaffiltext{10}{Santa Cruz Institute for Particle Physics,
University of California, Santa Cruz, CA 94064}
\altaffiltext{11}{Space Telescope Science Institute, 3700 San Martin
Drive, Baltimore, MD 21218}
\altaffiltext{12}{Department of Physics and Astronomy, 
University Of Waterloo, Waterloo, Ontario, Canada N2L 3G1}
\altaffiltext{13}{Department of Astronomy and Astrophysics, University
of Chicago, Chicago, IL 60637}
\altaffiltext{14}{Department of Physics, Stockholm University,
Albanova University Center, SE-106 91, Stockholm, Sweden}
\altaffiltext{15}{Subaru Telescope, National Astronomical Observatory
of Japan, 650 North A'ohaku Place, Hilo, HI 96720}
\altaffiltext{16}{National Astronomical Observatory of Japan, 2-21-1
Osawa, Mitaka, Tokyo,181-8588, Japan}
\altaffiltext{17}{Kavli Institute for Cosmological Physics, The University of
Chicago, Chicago IL 60637}
\altaffiltext{18}{Institute for Cosmic Ray Research, University of
Tokyo, 5-1-5, Kashiwanoha, Kashiwa, Chiba, 277-8582, Japan}
\altaffiltext{19}{Physikalisches Institut, Universit\"at Bonn, Bonn, Germany}
\altaffiltext{20}{Australian Astronomical Observatory, PO Box 296, Epping, 
NSW 1710, Australia}
\altaffiltext{21}{University of California, Davis, CA 95618}
\altaffiltext{22}{Department of Astronomy, Kyoto University, Sakyo-ku,
Kyoto 606-8502, Japan}
\altaffiltext{23}{ESO, Karl-Schwarzschild-Strasse 2, D-85748 Garching, Germany}
\altaffiltext{24}{Institute of Geophysics and Planetary Physics,
Lawrence Livermore National Laboratory, Livermore, CA 94550}
\altaffiltext{25}{JSPS Fellow}
\altaffiltext{26}{Deceased}
\begin{abstract}

We report a measurement of the Type Ia supernova (SN~Ia) rate in
galaxy clusters at $0.9<z<1.46$ from the \emph{Hubble Space Telescope
(HST)} Cluster Supernova Survey. This is the first cluster SN~Ia rate
measurement with detected $z>0.9$ SNe.  Finding $8 \pm 1$ cluster
SNe~Ia, we determine a SN~Ia rate of $0.50^{+0.23}_{-0.19}$ (stat)
$^{+0.10}_{-0.09}$ (sys) $h_{70}^2$ SNuB (SNuB $\equiv 10^{-12}$
SNe~$L_{\odot,B}^{-1}$~yr$^{-1}$). In units of stellar mass, this
translates to $0.36^{+0.16}_{-0.13}$ (stat) $^{+0.07}_{-0.06}$ (sys)
$h_{70}^2$ SNuM (SNuM $\equiv 10^{-12}$
SNe~$M_\odot^{-1}$~yr$^{-1}$). This represents a factor of $\approx
5 \pm 2$ increase over measurements of the cluster rate at $z<0.2$. We
parameterize the late-time SN~Ia delay time distribution with a power
law: $\Psi(t) \propto t^s$. Under the approximation of a single-burst
cluster formation redshift of $z_f = 3$, our rate measurement in
combination with lower-redshift cluster SN~Ia rates constrains $s =
-1.41^{+0.47}_{-0.40}$, consistent with measurements of the delay time
distribution in the field. This measurement is generally consistent
with expectations for the ``double degenerate'' scenario and
inconsistent with some models for the ``single degenerate'' scenario
predicting a steeper delay time distribution at large delay times.  We
check for environmental dependence and the influence of younger
stellar populations by calculating the rate specifically in cluster
red-sequence galaxies and in morphologically early-type galaxies,
finding results similar to the full cluster rate. Finally, the upper
limit of one host-less cluster SN~Ia detected in the survey implies
that the fraction of stars in the intra-cluster medium is less than
0.47 ($95\%$ confidence), consistent with measurements at lower
redshifts.

\end{abstract}
\section{Introduction} \label{intro}

Type Ia supernovae (SNe~Ia) are widely accepted to be the result of
the thermonuclear explosion of a carbon-oxygen (CO) white dwarf (WD).
The explosion is believed to occur as the WD nears the Chandrasekhar
mass by accreting mass from its companion star in a binary
system. Despite the confidence in this basic model, many uncertainties
remain about the process that leads to SNe~Ia \citep[see][for a
review]{livio01a}. Chief amongst them is the nature of the companion
donor star. The leading models fall into two classes: the {\it single
degenerate} scenario \citep[SD;][]{whelan73a}, and the {\it double
degenerate} scenario \citep[DD;][]{iben84a,webbink84a}. In the SD
scenario the companion is a red giant or main sequence star that
overflows its Roche lobe. In the DD scenario, the companion is a
second WD which merges with the primary after orbital decay due to the
emission of gravitational radiation. 

A better understanding of the SN~Ia progenitor is demanded from both
an astrophysical and a cosmological perspective. Astrophysically,
SNe~Ia dominate the production of
iron \citep[e.g.,][]{matteucci86a,tsujimoto95a,thielemann96a} and
provide energy feedback \citep{scannapieco06a} in galaxies. Knowledge
of the SN~Ia rate is necessary to include these effects in
galaxy evolution models. However, an accurate prediction of the SN~Ia
rate in galaxies of varying ages, masses and star formation histories
requires a good understanding of the nature of the progenitor. This is
particularly true for higher redshifts where direct SN rate
constraints are unavailable.  From a cosmological perspective, the
progenitor has become a central concern following the use of SNe~Ia as
standardizable candles in the discovery of dark
energy \citep{riess98a,perlmutter99a}. With hundreds of SNe now being
used in the precision measurement of cosmological
parameters \citep[e.g.,][]{hicken09a,amanullah10a}, astrophysical
sources of systematic error will soon become significant.  While the
unknown nature of the SN progenitor system is unlikely to bias
measurements at the current level of
uncertainty \citep{yungelson00a,sarkar08a}, it could become a
significant source of uncertainty in the future, as it leaves open the
question of whether high-redshift SNe are different than low-redshift
SNe in a way that affects the inferred distance.

Measuring the SN~Ia rate as a function of environment has long been
recognized as one of the few available methods for probing the SN~Ia
progenitor \citep[e.g.,][]{ruizlapuente95a,ruizlapuente98a,yungelson00a}.
SN~Ia rates constrain the progenitor scenario via the delay time
distribution (DTD), where ``delay time'' refers to the time between
star formation and SN~Ia explosion. The DTD is the distribution of
these times for a population of stars, and is equivalent to the SN~Ia
rate as a function of time after a burst of star formation. The delay
time is governed by different physical mechanisms in the different
progenitor scenarios. For example, in the SD scenario, when the donor
is a red giant star the delay time is set by the time the companion
takes to evolve off the main sequence.  In the DD scenario, it is
dominated by the time the orbit takes to decay due to gravitational
radiation.  The result is that the shape of the DTD depends on the
progenitor scenario.

However, the interpretation of the DTD is complicated by its
dependence on other factors, not all of which are completely
understood.  These include the initial mass function (IMF) of the
stellar population, the distribution of initial separation and mass
ratio in binary systems, and the evolution of the binary through one
or more common envelope \citep[CE; see, e.g.,][]{yungelson05a}
phases. Theoretical delay time distributions were computed
analytically following the proposal of both the SD \citep{greggio83a}
and DD \citep{tornambe86a,tornambe89a} scenarios. Later, theoretical
DTDs were extended to include various subclasses of each model and a
wider range of parameters \citep{tutukov94a,
yungelson00a,matteucci01a,belczynski05a,greggio05a}. In various recent
numerical simulations, different plausible prescriptions for the
initial conditions and for the binary evolution have lead to widely
ranging DTDs, even within one
scenario \citep{hachisu08a,kobayashi09a,ruiter09a,mennekens10a}. A
measurement of the DTD then must constrain not only the relative
contribution of various progenitor scenarios, but also the initial
conditions and CE phase, which is particularly poorly
constrained. Still, most simulations show a difference in the
DTD shape between the SD and DD scenarios. In both scenarios, the SN
rate is greatest shortly after star formation and gradually decreases
with time. However, the SD scenario typically shows a strong drop off
in the SN rate at large delay times not seen in the DD
scenario \citep[but see][]{hachisu08a}.

The DTD can be measured empirically from the SN~Ia
rate in stellar populations of different ages. Measurements
correlating SN rate with host star formation rate or star formation
history have now confirmed that the delay time spans a wide range,
from less than 100 Myr \citep[e.g.,][]{aubourg08a} to many
Gyr \citep[e.g.,][]{schawinski09a}. Correlations with star formation
rates \citep{mannucci05a,mannucci06a,sullivan06a,pritchet08a} show
that SNe with progenitor ages $\lesssim$ a few hundred Myr comprise
perhaps $\sim$50\% of all SNe~Ia. Measurements as a function of
stellar age \citep{totani08a,brandt10a}, show that the rate declines
with delay time as expected.

It is more straightforward to extract the DTD in stellar populations
with a narrow range of ages (with a single burst of star formation
being the ideal). Galaxy clusters, which are dominated by early-type
galaxies, provide an ideal environment for constraining the shape of
the DTD at large delay times. Early-type galaxies are generally
expected to have formed early ($z \gtrsim 2$) with little star
formation since \citep{stanford98a,vandokkum01a}. Cluster early-type
galaxies in particular form even earlier than those in the field, with
most star formation occurring at $z \gtrsim
3$ \citep{thomas05a,sanchezblazquez06a,gobat08a}.  Measuring the
cluster SN~Ia rate over a range of redshifts from $z=0$ to $z>1$
provides a measurement of the SN~Ia rate at delay times from $\sim$2
to 11 Gyr. Obtaining an accurate rate at the highest-possible redshift
is crucial for constraining the shape of the late-time DTD: a larger
redshift range corresponds to a larger lever arm in delay time.

In addition to DTD constraints, there are also strong motivations for
measuring the cluster SN~Ia rate from a perspective of cluster
studies. SNe~Ia are an important source of iron in the intracluster
medium \citep[e.g.,][]{loewenstein06a}. Cluster SN rates constrain the
iron contribution from SNe and, paired with measured iron abundances,
can also constrain possible enrichment mechanisms \citep{maoz04a}. The
high-redshift cluster rate is particularly important: measurements
show that most of the intracluster iron was produced at high
redshift \citep{calura07a}. The poorly-constrained high-redshift
cluster rate is one of the largest sources of uncertainty in
constraining the metal-loss fraction from galaxies \citep{sivanandam09a}.

Cluster SNe~Ia can also be used to trace the
diffuse \emph{intracluster} stellar component. Intracluster stars,
bound to the cluster potential rather than individual galaxies, have
been found to account for anywhere from $5\%$ to $50\%$ of the stellar
mass in clusters \citep[e.g.,][]{ferguson98a,feldmeier98a,
gonzalez00a,feldmeier04a,lin04a,zibetti05a,gonzalez05a,krick06a,mihos05a}.
The use of SNe~Ia as tracers of this component was first demonstrated
by \citet{galyam03a} who found two likely host-less SNe~Ia out of a
total of seven cluster SNe~Ia in $0.06 < z < 0.19$ Abell
clusters. After correcting for the greater detection efficiency of host-less
SNe, they determined that on average, the intracluster medium
contained $20^{+20}_{-12}$\% of the total cluster stellar mass.  The
intrinsic faintness of the light from intracluster stars, combined
with $(1+z)^4$ surface brightness dimming, makes surface brightness
measurements impossible at redshifts much higher than $z=0.3$.  Type
Ia supernovae, which are detectable up to and beyond $z=1$, provide a
way to measure the intracluster stellar component and its possible
evolution with redshift.

The cluster SN~Ia rate has recently been measured at lower redshifts
($z>0.3$) in several studies \citep{sharon07a,mannucci08a,dilday10a},
and at intermediate redshift ($z\sim 0.6$)
by \citet{sharon10a}. However, at higher redshifts ($z \gtrsim 0.8$),
only weak constraints on the high-redshift cluster Ia rate exist,
based on 1--2 SNe~Ia at $z=0.83$ \citep{galyam02a}.  In this paper, we
calculate the SN~Ia rate in $0.9 < z < 1.46$ clusters observed in the
{\it HST} Cluster Supernova Survey. We address the host-less SN~Ia
fraction, and use our result to place constraints on the late-time DTD
in clusters. \citet[][hereafter Maoz10]{maoz10c} have already combined
our results with iron abundance measurements and rate measurements in
other environments to place even tighter constraints on the SN~Ia DTD.

This paper is organized as follows. In \S\ref{survey} we review the
survey, placing particular emphasis on the aspects relevant to the
rate calculation. In \S\ref{sne} we describe the selection of
supernova candidates used in this rate calculation and the
determination of supernova type for these candidates.  In \S\ref{ct}
we carry out efficiency studies to determine the detection efficiency
of our SN selection. In \S\ref{lum} we measure the luminosity of the
clusters based on data from the survey. In \S\ref{results} we present
results and characterize systematic errors.  We discuss
interpretations for the delay time distribution and conclude
in \S\ref{conclusions}.  Throughout the paper we use a cosmology
with $H_0=70$~km~s$^{-1}$~Mpc$^{-1}$, $\Omega_M=0.3$, $\Omega_\Lambda =
0.7$. Unless otherwise noted, magnitudes are in the Vega system.

This paper is one of a series of ten papers that report supernova
results from the \emph{HST} Cluster Supernova Survey (PI: Perlmutter,
\emph{HST} program GO-10496), a survey to discover and follow SNe~Ia 
in very distant clusters. Paper I \citep[][hereafter Dawson09]{dawson09a} describes the
survey strategy and discoveries. This work, Paper II, reports on the
SN~Ia rate in clusters. Paper III \citep[][hereafter
Meyers11]{meyers11a} addresses the properties of the galaxies that
host SNe~Ia. Paper IV \citep{ripoche11a} introduces a new technique to
calibrate the zeropoint of the NICMOS camera at low counts rates,
critical for placing NICMOS-observed SNe~Ia on the Hubble
diagram. Paper V \citep{suzuki11a} reports the SNe~Ia lightcurves and
cosmology from the \emph{HST} Cluster SN Survey program. Paper VI
\citep{barbary11b} reports on the volumetric field SN~Ia
rate. \citet{melbourne07a}, one of several unnumbered papers in the
series, present a Keck adaptive optics observation of a $z=1.31$ SN~Ia
in $H$-band. \citet{barbary09a} report the discovery of the
extraordinary luminous supernova, SN~SCP06F6. \citet{morokuma10a}
presents the spectroscopic follow-up observations for SN
candidates. Finally, Hsiao et al. (in preparation) develop techniques
to remove problematic artifacts remaining after the standard STScI
pipeline. A separate series of papers, ten to date, reports on
cluster studies from the survey:
\citet{hilton07a,eisenhardt08a,jee09a,hilton09a,huang09a,rosati09a,
santos09a,strazzullo10a,brodwin10a,jee11a}.

\section{The Survey} \label{survey}

\begin{deluxetable*}{lccccc}
\tablewidth{0pt}
\tabletypesize{\scriptsize}
\tablecaption{\label{table:clusters} Cluster positions and redshifts}
\tablehead{\colhead{ID} & \colhead{Cluster} & \colhead{Redshift} & 
\colhead{R.A. (J2000)} & \colhead{Decl. (J2000)} & \colhead{Discovery}}
\startdata
A &  XMMXCS J2215.9-1738  & 1.46 & $22^{\rm h}\,15^{\rm m}\,59^{\rm s}.0$ & $-17^{\circ}\,37'\,59''$ & X-ray \\
B &  XMMU J2205.8-0159    & 1.12 & $22^{\rm h}\,05^{\rm m}\,50^{\rm s}.6$ & $-01^{\circ}\,59'\,30''$ & X-ray \\
C &  XMMU J1229.4+0151    & 0.97 & $12^{\rm h}\,29^{\rm m}\,29^{\rm s}.2$ & $+01^{\circ}\,51'\,21''$ & X-ray \\
D &  RCS J0221.6-0347     & 1.02 & $02^{\rm h}\,21^{\rm m}\,42^{\rm s}.2$ & $-03^{\circ}\,21'\,52''$ & Optical \\
E &  WARP J1415.1+3612    & 1.03 & $14^{\rm h}\,15^{\rm m}\,11^{\rm s}.1$ & $+36^{\circ}\,12'\,03''$ & X-ray \\
F &  ISCS J1432.4+3332    & 1.11 & $14^{\rm h}\,32^{\rm m}\,28^{\rm s}.1$ & $+33^{\circ}\,33'\,00''$ & IR-Spitzer \\
G &  ISCS J1429.3+3437    & 1.26 & $14^{\rm h}\,29^{\rm m}\,17^{\rm s}.7$ & $+34^{\circ}\,37'\,18''$ & IR-Spitzer \\
H &  ISCS J1434.4+3426    & 1.24 & $14^{\rm h}\,34^{\rm m}\,28^{\rm s}.6$ & $+34^{\circ}\,26'\,22''$ & IR-Spitzer \\
I &  ISCS J1432.6+3436    & 1.34 & $14^{\rm h}\,32^{\rm m}\,38^{\rm s}.8$ & $+34^{\circ}\,36'\,36''$ & IR-Spitzer \\
J &  ISCS J1434.7+3519    & 1.37 & $14^{\rm h}\,34^{\rm m}\,46^{\rm s}.0$ & $+35^{\circ}\,19'\,36''$ & IR-Spitzer \\
K &  ISCS J1438.1+3414    & 1.41 & $14^{\rm h}\,38^{\rm m}\,08^{\rm s}.2$ & $+34^{\circ}\,14'\,13''$ & IR-Spitzer \\
L &  ISCS J1433.8+3325    & 1.37 & $14^{\rm h}\,33^{\rm m}\,51^{\rm s}.1$ & $+33^{\circ}\,25'\,50''$ & IR-Spitzer \\
M &  Cl J1604+4304        & 0.90 & $16^{\rm h}\,04^{\rm m}\,23^{\rm s}.8$ & $+43^{\circ}\,04'\,37''$ & Optical \\
N &  RCS J0220.9-0333     & 1.03 & $02^{\rm h}\,20^{\rm m}\,55^{\rm s}.5$ & $-03^{\circ}\,33'\,10''$ & Optical \\
P &  RCS J0337.8-2844     & 1.1\tablenotemark{a} & $03^{\rm h}\,37^{\rm m}\,51^{\rm s}.2$ & $-28^{\circ}\,44'\,58''$ & Optical \\
Q &  RCS J0439.6-2904     & 0.95 & $04^{\rm h}\,39^{\rm m}\,37^{\rm s}.6$ & $-29^{\circ}\,05'\,01''$ & Optical \\
R &  XLSS J0223.0-0436    & 1.22 & $02^{\rm h}\,23^{\rm m}\,03^{\rm s}.4$ & $-04^{\circ}\,36'\,14''$ & X-ray \\
S &  RCS J2156.7-0448     & 1.07 & $21^{\rm h}\,56^{\rm m}\,42^{\rm s}.2$ & $-04^{\circ}\,48'\,04''$ & Optical \\
T &  RCS J1511.0+0903     & 0.97 & $15^{\rm h}\,11^{\rm m}\,03^{\rm s}.5$ & $+09^{\circ}\,03'\,09''$ & Optical \\
U &  RCS J2345.4-3632     & 1.04 & $23^{\rm h}\,45^{\rm m}\,27^{\rm s}.2$ & $-36^{\circ}\,32'\,49''$ & Optical \\
V &  RCS J2319.8+0038     & 0.90 & $23^{\rm h}\,19^{\rm m}\,53^{\rm s}.4$ & $+00^{\circ}\,38'\,13''$ & Optical \\
W &  RX J0848.9+4452      & 1.26 & $08^{\rm h}\,48^{\rm m}\,56^{\rm s}.4$ & $+44^{\circ}\,52'\,00''$ & X-ray \\
X &  RDCS J0910+5422      & 1.10 & $09^{\rm h}\,10^{\rm m}\,45^{\rm s}.1$ & $+54^{\circ}\,22'\,07''$ & X-ray \\
Y &  RDCS J1252.9-2927    & 1.24 & $12^{\rm h}\,52^{\rm m}\,54^{\rm s}.4$ & $-29^{\circ}\,27'\,17''$ & X-ray \\
Z &  XMMU J2235.3-2557    & 1.39 & $22^{\rm h}\,35^{\rm m}\,20^{\rm s}.8$ & $-25^{\circ}\,57'\,39''$ & X-ray 

\enddata
\tablenotetext{a}{photometric redshift}
\tablerefs{A \citep{stanford06a,hilton07a}; 
B,C \citep{bohringer05a,santos09a}; D \citep[also known as RzCS 052;][]{andreon08b,andreon08c}; 
D, N, U (Gilbank et al. in prep); E \citep{perlman02a};
F \citep{elston06a}; G, I, J, L \citep{eisenhardt08a}; 
L (Brodwin et al. in prep; Stanford et al. in prep);
H \citep{brodwin06a}; K \citep{stanford05a}; M \citep{postman01a};
Q \citep{cain08a}; R \citep{andreon05a,bremer06a};
S \citep{hicks08a}; V \citep{gilbank08a}; W \citep{rosati99a};
X \citep{stanford02a}; Y \citep{rosati04a}; 
Z \citep{mullis05a,rosati09a}.}
\tablecomments{Cluster positions differ
slightly from those reported in Dawson09 due to the use of an updated
algorithm for determining cluster centers.}
\end{deluxetable*}

The details of the \emph{HST} Cluster SN Survey are described in Dawson09.
Here, we briefly summarize the survey and highlight the details
relevant to the rate calculation.  The survey targeted 25 massive
galaxy clusters in a rolling search between July 2005 and December
2006.  Clusters were selected from X-ray, optical and IR surveys and
cover the redshift range $0.9<z<1.46$. Twenty-four of the clusters
have spectroscopically confirmed redshifts and the remaining
cluster has a photometric redshift estimate.  Cluster positions,
redshifts and discovery methods are listed in
Table~\ref{table:clusters}. Note that cluster positions differ
slightly from those reported in Dawson09 due to the use of an updated
algorithm for determining cluster centers.

\begin{figure}
\epsscale{1.175}
\plotone{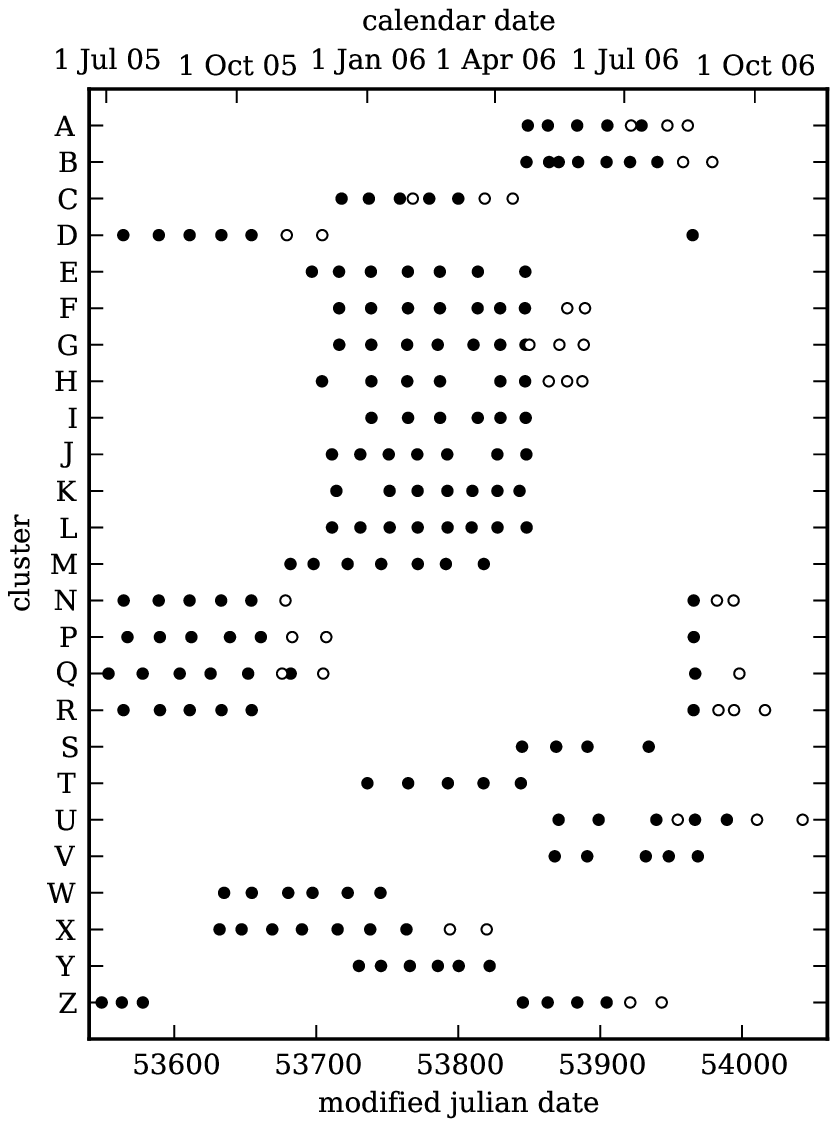}
\caption{Dates of visits to each cluster. All visits included $z_{850}$ 
exposures (usually four). Most visits also included one $i_{775}$
exposure. Filled circles indicate ``search'' visits (used for finding
SNe). Open circles indicate ``follow-up'' visits (contingent on the
existence of an active SN candidate). Clusters D, N, P, Q and R were
re-visited once towards the end of the survey, with additional
follow-up visits devoted to clusters in which promising SN candidates
were found (N, Q, R). \label{fig:visits}}
\end{figure}

During the survey, each cluster was observed once every 20 to 26 days
during its \emph{HST} visibility window (typically four to seven
months). Figure~\ref{fig:visits} shows the dates of visits to
each cluster. Each visit consisted of four exposures in the F850LP
filter (hereafter $z_{850}$).  Most visits also included a fifth
exposure in the F775W filter (hereafter $i_{775}$). We revisited
clusters D, N, P, Q, R and Z towards the end of the survey when they
became visible again.

Immediately following each visit, the four $z_{850}$ exposures were
cosmic ray-rejected and combined using {\sc MultiDrizzle}
\citep{fruchter02a,koekemoer02a} and searched for
supernovae. Following the technique employed in the earliest Supernova
Cosmology Project searches \citep{perlmutter95a,perlmutter97a}, we
used the initial visit as a reference image, flagged candidates with
software and then considered them by eye. Likely supernovae were
followed up spectroscopically using pre-scheduled time on the Keck,
and Subaru telescopes and target-of-opportunity observations on VLT.
For nearly all SN candidates, either a live SN spectrum or host galaxy
spectrum was obtained. In many cases, spectroscopy of cluster galaxies
was obtained contemporaneously using slit masks. Candidates deemed
likely to be at higher redshift ($z>1$) were also observed with the
NICMOS camera on {\it HST}, but these data are not used in this work.

A number of visits were contingent on the existence of an active
SN. At the end of a cluster's visibility window, the last two
scheduled visits were cancelled if there was no live SN previously
discovered. This is because a SN discovered on the rise in either of
the last two visits could not be followed long enough to obtain a
cosmologically useful light curve. In addition, supplementary visits
between pre-scheduled visits were occasionally added to provide more
complete light curve information for SNe (in the case of clusters A,
C, Q, and U). We call all visits contingent on the existence of an
active SN ``follow-up'' visits (designated by open circles in
Fig.~\ref{fig:visits}).

\section{Supernova Selection} \label{sne}

During the survey, our aim was to find as many supernovae as possible
and find them as early as possible in order to trigger spectroscopic
and NICMOS follow-up. Thus, software thresholds for flagging candidates
for consideration were set very low, and all possible supernovae were
carefully considered by a human screener. Over the course of the survey,
thresholds were changed and the roster of people scanning the
subtractions changed. As a result, the initial candidate selection
process was inclusive but heterogeneous, and depended heavily on human
selection. This made it difficult to calculate a selection efficiency
for the SN candidates selected during the survey (listed in Tables~3
and 4 of Dawson09).

In this section, we select an independent SN candidate sample (without
regard for the Dawson09 sample) using automated selection wherever
possible. Although the remainder of this paper will focus on cluster
SNe, candidates are selected without regard for cluster membership
(which is only known from follow-up spectroscopy once the candidate
has already been found) and we determine SN types for both cluster and
non-cluster SNe. The non-cluster SNe are considered further in a
second paper deriving the volumetric SN~Ia field rate (Barbary et al.,
in preparation).  The automated selection consists of initial
detection in pairs of subtracted images (\S\ref{search}; 86 candidates
selected), and subsequent requirements based on the light curve of
each candidate (\S\ref{lccuts}; 60 candidates remaining). The
selection efficiency for these two steps is later calculated via a
Monte Carlo simulation. In \S\ref{typing} we assign a type (SN~Ia,
core-collapse SN, or other) to each of the remaining 60 candidates
based on all data available (including triggered follow-up
observations). For this last step we do not calculate an efficiency or
completeness. Instead we estimate the classification uncertainty of
the assigned type for each candidate individually. For most candidates
the uncertainty in the type is negligible thanks to ample photometric
and spectroscopic data.

\subsection{Initial detection} \label{search}

\begin{figure*}
\begin{center}
\epsscale{1.15}
\plotone{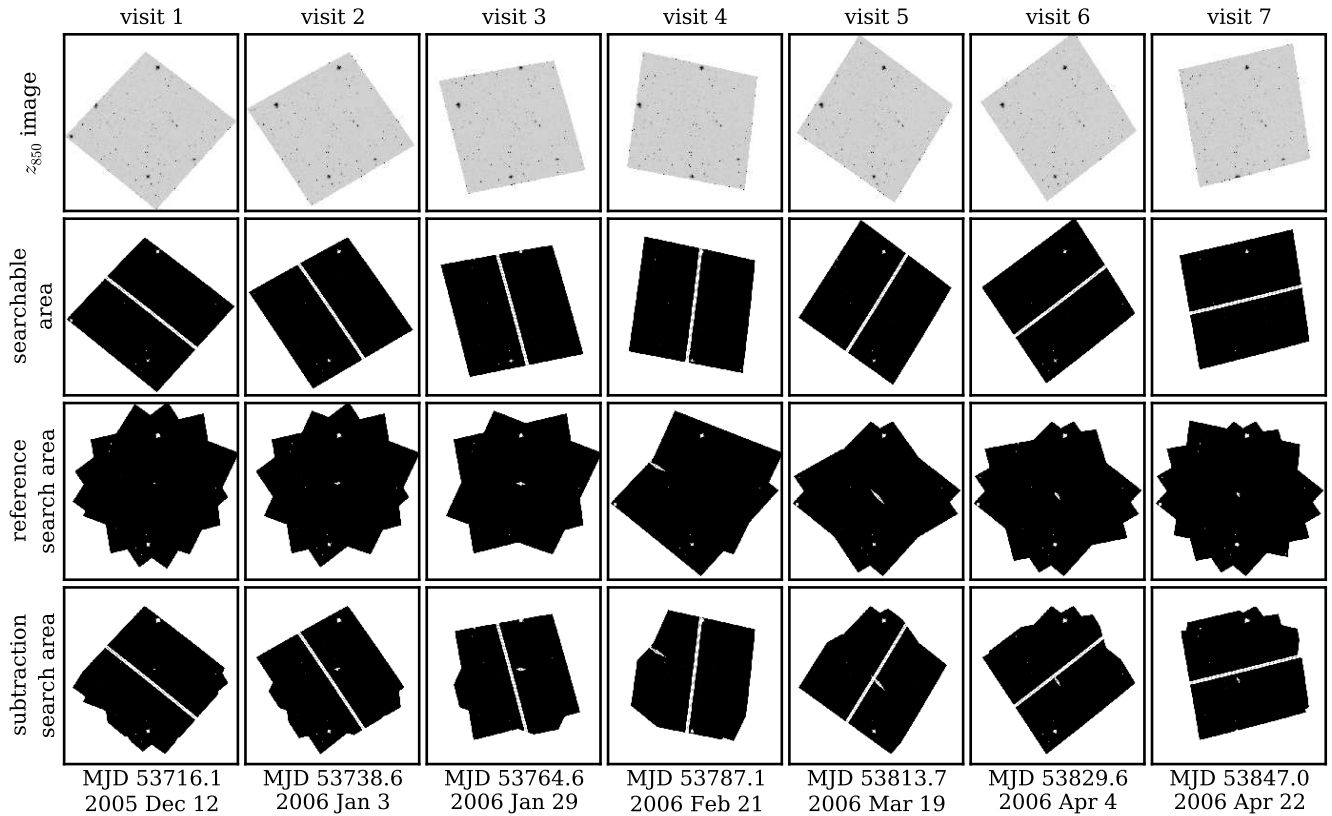}
\end{center}
\caption{An example of image orientation and searchable regions for 
         cluster ISCS J1432.4+3332. Each column represents an
         observation of the cluster. The first row is the $z_{850}$
         image for that visit. The second row is the part of that
         image that is searchable. The third row shows the searchable
         area of the stacked reference image used in the subtraction
         for this visit. The fourth row is the searchable area in the
         subtraction (the intersection of the second and third rows).
         \label{fig:epochs}}
\end{figure*}

For the purpose of initially detecting candidates, we use only
``search'' visits (filled circles in Fig.~\ref{fig:visits}) and
disregard the ``follow-up'' visits (open circles in
Fig.~\ref{fig:visits}). (In the following section we will use any
available ``follow-up'' visits to construct more complete light curves
for the candidates discovered in this section.) We use the {\sc
MultiDrizzle}-combined, cosmic ray-rejected, $z_{850}$ image from each
``search'' visit. We consider only regions in this image that are
covered by three or more $z_{850}$ exposures.  With less than three
exposures, the combined images are too heavily contaminated by cosmic
rays to be practically searchable for SNe. Although there are
typically four $z_{850}$ exposures, the dither pattern used in the
survey means that not all regions of the combined image have four
exposures. The ACS camera is a mosaic of two $2048 \times 4096$~pixel
CCD chips (1~pixel = $0.05''$) separated by $2.5''$. The $z_{850}$
exposures were dithered to cover this gap, meaning that a $5''$ wide
region in the center of the image and $2.5''$ wide regions on either
side of the image are only covered by two exposures and thus are not
searchable. Due to orbital constraints, the position angle of {\it HST}
changes between each visit. This means that the unsearchable ``gap''
region rotates over the field between visits, and that the outer parts
of the field are observed in some visits, but not others
(Fig.~\ref{fig:epochs}, second row). The regions around bright stars
are also considered ``not searchable'' and are similarly masked.

For each ``search'' visit to each cluster, we follow these four steps:

{\bf 1. A reference image is made} by combining images from
other visits to the cluster. All visits that are either 50 or more
days before the search epoch or 80 or more days after the search epoch
are included. If there are no epochs outside this 130~day range, the
range is narrowed symmetrically until one epoch qualifies. Masked
pixels in each visit's image do not contribute to the stacked
reference image (Fig.~\ref{fig:epochs}, third row).

{\bf 2. A subtracted image is made} by subtracting the stacked
reference image from the search epoch image. A map of the sky noise
level in the subtraction is made by considering the noise level of the
search epoch image and the noise level of each reference image
contributing to a given region. Any area masked in either the search
epoch or stacked reference image is masked in the subtracted image
(Fig.~\ref{fig:epochs}, fourth row).

{\bf 3. Candidates in the subtraction are identified by software.} To
  be flagged, a candidate must have three contiguous pixels with a flux
  3.4 times the local sky noise level in the subtraction (as
  determined by the sky noise map above). Once flagged, it must
  fulfill the following four requirements:
\begin{itemize}
\item {\sc MultiDrizzle}-combined image: A total signal-to-noise ratio 
  (including sky and Poisson noise) of 5 or more in a 3~pixel radius
  aperture.
\item {\sc MultiDrizzle}-combined image: A total signal-to-noise ratio 
  of 1.5 or more in a 10~pixel radius aperture.
\item Individual exposures: A signal-to-noise ratio of 1 or greater 
  in a 3~pixel radius aperture in three or more individual exposures.
\item Individual exposures: A candidate cannot have an individual exposure 
  with a flux more than $20\sigma$ greater than the flux in the lowest 
  flux exposure \emph{and} a second individual exposure with flux more 
  than $10\sigma$ greater than the flux in the lowest flux exposure.
\end{itemize}
The first requirement is designed to eliminate low significance
detections on bright galaxies. The second requirement helps eliminate
dipoles on bright galaxy cores caused by slight image misalignment. The
third and fourth requirements are aimed at false detections due to
cosmic ray coincidence. They require the candidate to be detected in
most of the exposures and allow no more than one exposure to be
greatly affected by a cosmic ray. On the order of five to ten
candidates per subtraction pass all the requirements, resulting in
approximately 1000 candidates automatically flagged across the 155
search visits.

{\bf 4. Each candidate is evaluated by eye in the subtraction.}
Because the position angle changes between each epoch, the orientation
of stellar diffraction spikes changes, causing the majority of the
false detections. These are easy to detect and eliminate by
eye. Occasionally there are mis-subtractions on the cores of bright
galaxies that pass the above requirements. Only completely unambiguous
false detections are eliminated in this step. If there is any
possibility the candidate is a real SN, it is left in the sample for
further consideration.

After carrying out the above four steps for all 155 search visit, 86
candidates remain. At this point, candidates have been selected based
only on information from a single $z_{850}$ subtraction. Detailed information on each of the 86 candidates is available from the \emph{HST} Cluster SN Survey website\footnote{\url{http://supernova.lbl.gov/2009ClusterSurvey}}.

\subsection{Lightcurve Requirements} \label{lccuts}
The 86 remaining candidates still include a considerable number of
non-SNe. We wish to trim the sample down as much as possible in an
automated way, so that we can easily calculate the efficiency of our
selection.  For each candidate, we now make three further automated
requirements based on $i_{775}$ data (if available) and the shape of
the $z_{850}$ light curve.  The requirements and number of candidates
remaining after each requirement are summarized in
Table~\ref{table:lccuts}.

\begin{deluxetable}{lc}

\tablewidth{0pt}
\tablecaption{\label{table:lccuts} Light Curve Requirements}
\tablehead{\colhead{Requirement} & \colhead{Candidates Remaining}}
\startdata
Before light curve requirements                   & 86\\
Positive $i_{775}$ flux (if observed in $i_{775}$) & 81\\
$2\sigma$ Detection in surrounding epochs         & 73\\
If declining, Require two $5\sigma$ detections    & 60
\enddata

\end{deluxetable}

First, we require that if $i_{775}$ data exists for the epoch in which
the candidate was detected, there be positive flux in a 2~pixel radius
aperture at the candidate location in the $i_{775}$ image. From our SN
light curve simulations, we find that virtually all SNe should pass
(near maximum light there is typically enough SN flux in the $i_{775}$
filter to result in a positive total flux, even with large negative
sky fluctuations).  Meanwhile, about half of the cosmic rays located
far from galaxies will fail this test (due to negative sky
fluctuations). If there is no $i_{775}$ data for the detection epoch,
this requirement is not applied. Even though nearly all SNe are
expected to pass, we account for any real SNe that would be removed in
our Monte Carlo simulation.

Second, we require that the light curve does not rise and fall too
quickly: if there is a ``search'' visit less than 60~days before the
detection visit and also one less than 60~days after the detection visit,
the candidate must be detected at a $2\sigma$ level in at least one of
these two visits. SNe~Ia have light curves wide enough to be detected
at this level in two epochs spaced apart by 60~days. However, cosmic
rays in one $z_{850}$ image are unlikely to be repeated in the same
spot in two epochs and thus will be removed. This requirement is also
included in our Monte Carlo simulation.

The third and final requirement aims to eliminate candidates that were
significantly detected in only the first epoch and that then faded
from view. Such candidates would not have been followed up
spectroscopically and it would typically be impossible to tell if such
candidates were SNe (and if so, Type~Ia or core collapse) on the basis
of a single detection. We chose to eliminate any such candidates and
account for this elimination in our Monte Carlo simulation, rather
than dealing with an ``untypeable'' candidate.  Specifically, if a
candidate is found on the decline (in the first search epoch), we
require two epochs with $5\sigma$ detections.  For high-redshift
($z \sim 1$) SNe~Ia, this requirement means that the first epoch will
be at approximately maximum light, and most of the SN decline is
captured, making it possible to confirm a SN and estimate a type.
For candidates that are only significantly detected in the last search
epoch, typing is not a problem because additional ACS orbits were
typically scheduled in order to follow such candidates.

After these requirements 60 candidates remain. The automatic selection
means that we can easily calculate the completeness of the selection
so far; any real SNe~Ia removed will be accounted for in the
``effective visibility time'' (\S\ref{ct}) which is calculated using a
Monte Carlo simulation.

\subsection{Typing} \label{typing}

We now use all available information about each candidate
(spectroscopic confirmation, host galaxy redshift, all light curve
information, as well as host galaxy luminosity and color) to classify
each of the 60 remaining candidates as image artifact, active galactic
nucleus (AGN), core-collapse SN (SN~CC), or SN~Ia.

\vskip 0.3in
\subsubsection{Image artifacts}

Although the automated selections were designed to eliminate image
artifacts such as subtraction residuals and cosmic rays, they were
made to be somewhat tolerant so that real SNe were not eliminated. The
result is that some artifacts slip through. Candidates located close
to the cores of relatively bright galaxies that show adjoining
negative and positive areas in subtractions are likely to be caused by
mis-alignment between the reference and search image. For such
candidates, we inspected the full light curve for consistency with the
general shape of a SN~Ia light curve. For fourteen of these, the
light curve is completely inconsistent with that of a SN~Ia. Their
light curves have either multiple peaks, long flat portions followed
by one or two lower points, and/or $i_{775}$ data that shows no
change. We classify these fourteen candidates as subtraction residuals
with negligible classification uncertainty (very unlikely that any are
SNe~Ia).
 
Candidates where one or two of the four $z_{850}$ exposures was
clearly affected by a cosmic ray or hot pixel may be false
detections. These can pass the automated cosmic ray rejection when
they occur on a galaxy.  For two such candidates, we used the lack of
any change in the $i_{775}$ light curve to rule out a SN~Ia: fitting
SN templates with a range of redshifts and extinctions resulted in
observed $i_{775}$ fluxes too low by $4\sigma$ or more, given the
$z_{850}$ increase. One other candidate, SCP06W50, is less certain. It
was discovered in the last visit to the cluster, making it difficult
to constrain a template light curve. There is clearly a hot pixel or
cosmic ray in one $z_{850}$ exposure, but there appears to be some
excess flux in the other three exposures as well. Also, there is a
point-source like detection in $i_{775}$, but offset $\sim$1.2~pixels
from the $z_{850}$ detection. While the $i_{775}$ detection may also
be a cosmic ray, it is possible that this candidate is a SN caught
very early. The elliptical ``host'' galaxy was not observed
spectroscopically, but we estimate its redshift to be $0.60 < z < 0.85$
based on the color of $i_{775} - z_{850} = 0.55$ and stellar
population models of \citet[][hereafter BC03]{bruzual03a}.

Of the 17 total candidates classified as image artifacts, SCP06W50 is the
only one with significant uncertainty. However, this uncertainty does
not affect the cluster SN~Ia rate as the host galaxy is clearly in the
cluster foreground.

\subsubsection{AGN}

\begin{figure*}
\epsscale{1.175}
\plotone{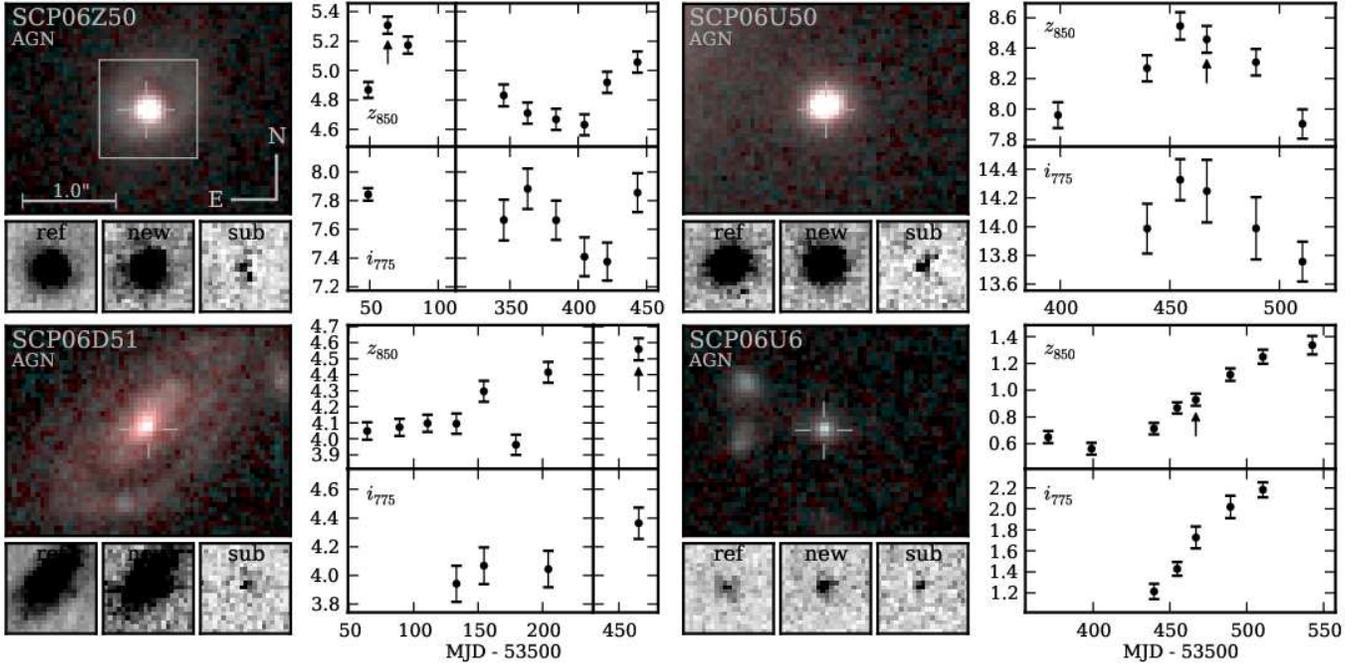}
\caption{Images and light curves of four of the 14 candidates 
classified as AGN. For each candidate,
the upper left panel shows the two-color stacked image ($i_{775}$ and
$z_{850}$) of the host galaxy, with the position of the transient
indicated.  The three smaller panels below the
stacked image show the reference, new, and subtracted images for the
discovery visit. The right panel shows the light curve at the SN
position (including host galaxy light) in the $z_{850}$ ({\it top})
and $i_{775}$ ({\it bottom}) bands. The y axes have units of counts
per second in a $3$~pixel radius aperture. The effective zeropoints
are 23.94 and 25.02 for $z_{850}$ and $i_{775}$, respectively. The
discovery visit is indicated with an arrow in the $z_{850}$ plot.
\label{fig:agn}}
\end{figure*}

Candidates positioned directly on the cores of their host galaxies may
be AGN. Four such candidates were spectroscopically confirmed as AGN:
SCP06L22 ($z=1.369$), SCP06V6 ($z=0.903$) and SCP05X13 ($z=1.642$) and
SCP06U3 ($z=1.534$). A fifth candidate, SCP06F3, is spectroscopically
consistent with an AGN at $z=1.21$, but is less
certain \citep[see spectroscopy reported in][]{morokuma10a}. SCP06L22,
SCP05X13, SCP06U3 and SCP06F3 also have light curves that are clearly
inconsistent with SNe~Ia (observer frame rise times of 100~days or
more, or declining phases preceding rising phases). Of the ``on core''
candidates that were not observed spectroscopically, five exhibit
light curves that decline before rising or have rise times of 100~days
or more. A sixth candidate, SCP06Z51 exhibited slightly varying fluxes
that could be due to either subtraction residuals or an AGN.  However,
its light curve is clearly inconsistent with a SN~Ia, especially
considering the apparent size, magnitude and color of the host
galaxy. Summarizing, there are 11 ``on-core'' candidates certain not
to be SNe~Ia.

Three other ``on-core'' candidates are also considered likely AGN on
the basis of their light curves: SCP06Z50, SCP06U50 and
SCP06D51. These three candidates are shown in
Figure~\ref{fig:agn}. SCP06Z50 (Fig.~\ref{fig:agn}, top left), has a
rise-fall behavior in the first three $z_{850}$ observations of its
light curve that \emph{could} be consistent with a SN~Ia light
curve. However, given that the host galaxy is likely at $z \lesssim 1$
based on its magnitude and color, the SN would be fainter than a
normal SN~Ia by 1~magnitude or more. Considering the proximity to the
galaxy core and the additional variability seen in the last two
observations, SCP06Z50 is most likely an AGN. The light curve of
candidate SCP06U50 (Fig.~\ref{fig:agn}, top right) also exhibits a
rise-fall that could be consistent with a supernova light
curve. However, its host is morphologically elliptical and likely at
$z \lesssim 0.7$ based on its color. At $z \lesssim 0.7$, a SN~Ia
would have to be very reddened ($E(B-V) \gtrsim 1$) to match the color
and magnitude of the SCP06U50 light curve. As this is very unlikely
(considering that the elliptical host likely contains little dust), we
conclude that SCP06U50 is also most likely an AGN. Finally, SCP06D51
(Fig.~\ref{fig:agn}, bottom left) was discovered in the last visit, on
the core of a spiral galaxy. We classify it as an AGN based on the
earlier variability in the light curve. As these galaxies are all most
likely in the cluster foregrounds, even the small uncertainty in these
classifications is not a concern for the cluster rate calculation
here.

Note that one of the candidates classified here as a clear AGN, SCP06U6,
was reported as a SN with unknown redshift by Dawson09, due to the fact
that spectroscopy revealed no evidence of an AGN.  However, it is on
the core of a compact galaxy, and has a clear $\gtrsim 100$ day rise
in both $z_{850}$ and $i_{775}$ (Fig.~\ref{fig:agn}, bottom
right). While it could possibly be a very peculiar SN with a long rise
time, what is important for this analysis is that it is clearly
not a SN~Ia.

\subsubsection{Supernovae}

After removing 17 image artifacts and 14 AGN, 29 candidates remain
(listed in Table~\ref{table:sn}). One of these is the peculiar
transient SCP~06F6 (also known as SN~SCP06F6) reported
by \citet{barbary09a}. Various explanations have been considered by,
e.g., \citet{gansicke09a}, \citet{soker10a}
and \citet{chatzopoulos09a}. It appears that SCP~06F6 may be a rare
type of supernova, with redshift
$z=1.189$ \citep{quimby09a,pastorello10a}. While its precise
explanation is still uncertain, the important fact for this analysis is
that SCP~06F6 is clearly not a SN~Ia.

\begin{deluxetable*}{lccccccc}

\tablecaption{\label{table:sn} Supernovae}

\tablehead{\colhead{ID} & \colhead{Nickname} & \colhead{R.A. (J2000)} & \colhead{Decl. (J2000)} & \colhead{$z$} & \colhead{SN Type} & \colhead{Confidence} & \colhead{Typing}}
\startdata
\sidehead{\emph{Cluster Members}}
SN SCP06C1 & Midge & $12^{\rm h}\,29^{\rm m}\,33^{\rm s}.012$ & $+01^{\circ}\,51'\,36''.67$ & 0.98\phn & Ia & secure & a,c\\
SN SCP05D0 & Frida & $02^{\rm h}\,21^{\rm m}\,42^{\rm s}.066$ & $-03^{\circ}\,21'\,53''.12$ & 1.014 & Ia & secure & a,b,c\\
SN SCP06F12 & Caleb & $14^{\rm h}\,32^{\rm m}\,28^{\rm s}.748$ & $+33^{\circ}\,32'\,10''.05$ & 1.11\phn & Ia & probable & c\\
SN SCP06H5 & Emma & $14^{\rm h}\,34^{\rm m}\,30^{\rm s}.139$ & $+34^{\circ}\,26'\,57''.29$ & 1.231 & Ia & secure & b,c\\
SN SCP06K18 & Alexander & $14^{\rm h}\,38^{\rm m}\,10^{\rm s}.663$ & $+34^{\circ}\,12'\,47''.19$ & 1.412 & Ia & probable & b\\
SN SCP06K0 & Tomo & $14^{\rm h}\,38^{\rm m}\,08^{\rm s}.366$ & $+34^{\circ}\,14'\,18''.08$ & 1.416 & Ia & secure & b,c\\
SN SCP06R12 & Jennie & $02^{\rm h}\,23^{\rm m}\,00^{\rm s}.082$ & $-04^{\circ}\,36'\,03''.04$ & 1.212 & Ia & secure & b,c\\
SN SCP06U4 & Julia & $23^{\rm h}\,45^{\rm m}\,29^{\rm s}.429$ & $-36^{\circ}\,32'\,45''.73$ & 1.05\phn & Ia & secure & a,c\\
\sidehead{\emph{Cluster Membership Uncertain}}
SN SCP06E12 & Ashley & $14^{\rm h}\,15^{\rm m}\,08^{\rm s}.141$ & $+36^{\circ}\,12'\,42''.94$ & \nodata & Ia & plausible & c\\
SN SCP06N32 & \nodata & $02^{\rm h}\,20^{\rm m}\,52^{\rm s}.368$ & $-03^{\circ}\,34'\,13''.32$ & \nodata & CC & plausible & c\\
\sidehead{\emph{Not Cluster Members}}
SN SCP06A4 & Aki & $22^{\rm h}\,16^{\rm m}\,01^{\rm s}.077$ & $-17^{\circ}\,37'\,22''.09$ & 1.193 & Ia & probable & c\\
SN SCP06B3 & Isabella & $22^{\rm h}\,05^{\rm m}\,50^{\rm s}.402$ & $-01^{\circ}\,59'\,13''.34$ & 0.743 & CC & probable & c\\
SN SCP06C0 & Noa & $12^{\rm h}\,29^{\rm m}\,25^{\rm s}.654$ & $+01^{\circ}\,50'\,56''.58$ & 1.092 & Ia & secure & b,c\\
SN SCP06C7 & \nodata & $12^{\rm h}\,29^{\rm m}\,36^{\rm s}.517$ & $+01^{\circ}\,52'\,31''.47$ & 0.61\phn & CC & probable & c\\
SN SCP05D6 & Maggie & $02^{\rm h}\,21^{\rm m}\,46^{\rm s}.484$ & $-03^{\circ}\,22'\,56''.18$ & 1.314 & Ia & secure & b,c\\
SN SCP06F6 & \nodata & $14^{\rm h}\,32^{\rm m}\,27^{\rm s}.394$ & $+33^{\circ}\,32'\,24''.83$ & 1.189 & non-Ia & secure & a\\
SN SCP06F8 & Ayako & $14^{\rm h}\,32^{\rm m}\,24^{\rm s}.525$ & $+33^{\circ}\,33'\,50''.75$ & 0.789 & CC & probable & c\\
SN SCP06G3 & Brian & $14^{\rm h}\,29^{\rm m}\,28^{\rm s}.430$ & $+34^{\circ}\,37'\,23''.13$ & 0.962 & Ia & plausible & c\\
SN SCP06G4 & Shaya & $14^{\rm h}\,29^{\rm m}\,18^{\rm s}.743$ & $+34^{\circ}\,38'\,37''.38$ & 1.35\phn & Ia & secure & a,b,c\\
SN SCP06H3 & Elizabeth & $14^{\rm h}\,34^{\rm m}\,28^{\rm s}.879$ & $+34^{\circ}\,27'\,26''.61$ & 0.85\phn & Ia & secure & a,c\\
SN SCP06L21 & \nodata & $14^{\rm h}\,33^{\rm m}\,58^{\rm s}.990$ & $+33^{\circ}\,25'\,04''.21$ & \nodata & CC & plausible & c\\
SN SCP06M50 & \nodata & $16^{\rm h}\,04^{\rm m}\,25^{\rm s}.300$ & $+43^{\circ}\,04'\,51''.85$ & \nodata & \nodata & \nodata & \nodata\\
SN SCP05N10 & Tobias & $02^{\rm h}\,20^{\rm m}\,52^{\rm s}.878$ & $-03^{\circ}\,33'\,40''.20$ & 0.203 & CC & plausible & c\\
SN SCP06N33 & Naima & $02^{\rm h}\,20^{\rm m}\,57^{\rm s}.699$ & $-03^{\circ}\,33'\,23''.97$ & 1.188 & Ia & probable & c\\
SN SCP05P1 & Gabe & $03^{\rm h}\,37^{\rm m}\,50^{\rm s}.352$ & $-28^{\circ}\,43'\,02''.66$ & 0.926 & Ia & probable & c\\
SN SCP05P9 & Lauren & $03^{\rm h}\,37^{\rm m}\,44^{\rm s}.512$ & $-28^{\circ}\,43'\,54''.58$ & 0.821 & Ia & secure & a,c\\
SN SCP06U7 & Ingvar & $23^{\rm h}\,45^{\rm m}\,33^{\rm s}.867$ & $-36^{\circ}\,32'\,43''.48$ & 0.892 & CC & probable & c\\
SN SCP06X26 & Joe & $09^{\rm h}\,10^{\rm m}\,37^{\rm s}.889$ & $+54^{\circ}\,22'\,29''.07$ & 1.44\phn & Ia & plausible & c\\
SN SCP06Z5 & Adrian & $22^{\rm h}\,35^{\rm m}\,24^{\rm s}.966$ & $-25^{\circ}\,57'\,09''.61$ & 0.623 & Ia & secure & a,c
\enddata

\tablecomments{Typing: (a) Spectroscopic confirmation. 
(b) Host is morphologically early-type, with no signs of recent star
formation.  (c) Light curve shape, color, magnitude consistent with
type. We do not assign a type for SCP06M50 because there is
significant uncertainty that the candidate is a SN at all.}

\end{deluxetable*}

Note that Table~\ref{table:sn} contains 10 fewer candidates than the
list presented by Dawson09. This is unsurprising; here we have
intentionally used a stricter selection than in the original search,
the source for the Dawson09 sample. Still, after finalizing our
selection method we checked that there were no unexpected
discrepancies. Five of the Dawson09 candidates (SCP06B4, SCP06U2,
SCP06X18, SCP06Q31, SCP06T1) fell just below either the detection or
signal-to-noise thresholds in our selection. These were found in the
original search because detection thresholds were set slightly lower,
and because the images were sometimes searched in several different
ways. For example, in the original search SCP06B4 was only found by
searching an $i_{775}$ subtraction. Two Dawson09 candidates (SCP05D55,
SCP06Z52) were found too far on the decline and failed the light curve
requirements (\S\ref{lccuts}). Three Dawson09 candidates (SCP06X27,
SCP06Z13, SCP06Z53) were found while searching in ``follow-up''
visits, which were not searched here. SCP06U6 passed all requirements,
but is classified here as an AGN, as noted above. With the exception
of SCP06U6, all of these candidates are likely to be supernovae
(mostly core collapse). However, the types of candidates that did not
pass our requirements are not of concern for this analysis. Finally,
SCP06M50 was not reported in Dawson09, but is classified here as a SN,
although a highly uncertain one (discussed in detail
in \S\ref{sncomments}).

Thanks to the extensive ground-based spectroscopic follow-up campaign,
we were able to obtain spectroscopic redshifts for 25 of the 29
SNe. The redshift reported in Table~\ref{table:sn} is derived from the
SN host galaxy for all but one candidate (SCP06C1) where the redshift
is from the SN spectrum itself. Of the 25 candidates with redshifts,
eight are in clusters and 17 are in the field. Note that this high
spectroscopic completeness is particularly important for determining
the cluster or non-cluster status of each SN, which directly affects
the determination of the cluster SN~Ia rate. The possible cluster
memberships of the four candidates lacking redshifts are discussed
below.

We determine the type of each of the 29 supernovae using a combination
of methods in order to take into account all available information for
each supernova. This includes (a) spectroscopic confirmation, (b) the
host galaxy environment, and (c) the SN light curve. To qualify the
confidence of each supernova's type, we rank the type as ``secure,''
``probable,'' or ``plausible'':
\begin{description}
\item[Secure SN~Ia] Has
spectroscopic confirmation or \emph{both} of the following: (1) an
early-type host galaxy with no recent star formation and (2) a light
curve with shape, color and magnitude consistent with SNe~Ia and
inconsistent with other types.
\item[Probable SN~Ia] Fulfills either the host galaxy 
requirement or the light curve requirement, but not both.
\item[Plausible SN~Ia] The light curve is indicative of a SN~Ia, 
but there is not enough data to rule out other types.
\item[Secure SN~CC] Has spectroscopic confirmation (note that 
there are no such candidates in this sample). 
\item[Probable SN~CC] The light curve is consistent with a 
core-collapse SN and inconsistent with a SN~Ia. 
\item[Plausible SN~CC] Has a light curve indicative
of a core-collapse SN, but not inconsistent with a SN~Ia. 
\end{description}
This ranking system is largely comparable to the ``gold,'' ``silver,''
``bronze'' ranking system of \citet{strolger04a}, except that we do
not use their ``UV deficit'' criterion. This is because our data do
not include the bluer F606W filter, and because SNe~Ia and CC are
only distinct in UV flux for a relatively small window early in the
light curve. Below, we discuss in detail the three typing methods used.

{\bf (a) Spectroscopic confirmation:} During the survey, seven
candidates were spectroscopically confirmed as
SNe~Ia \citep[Dawson09,][]{morokuma10a}. These seven (three of which are in
clusters) are designated with an ``a'' in the ``typing'' column of
Table~\ref{table:sn}. All seven candidates have a light curve shape,
absolute magnitude and color consistent with a SN~Ia. Although the
spectroscopic typing by itself has some degree of uncertainty, the
corroborating evidence from the light curve makes these ``secure''
SNe~Ia.

{\bf (b) Early-type host galaxy:} The progenitors of core-collapse SNe
are massive stars ($> 8 M_{\odot}$) with main sequence lifetimes of
$<40$~Myr. Thus, core-collapse SNe only occur in galaxies with recent
star formation. Early-type galaxies, having typically long ceased star
formation, overwhelmingly host Type~Ia
SNe \citep[e.g.,][]{cappellaro99a,hamuy00a}. In fact, in an extensive
literature survey of core-collapse SNe reported in early-type
hosts, \citet{hakobyan08a} found that only three core-collapse SNe
have been recorded in early-type hosts, and that the three host
galaxies in question had either undergone a recent merger or were
actively interacting. In all three cases there are independent
indicators of recent star formation. Therefore, in the cases where the
host galaxy morphology, photometric color, and spectrum all indicate
an early-type galaxy with no signs of recent star formation or
interaction, we can be extremely confident that the SN type is
Ia. These cases are designated by a ``b'' in the ``typing'' column of
Table~\ref{table:sn}. We emphasize that in all of these cases,
spectroscopy reveals no signs of recent star formation and there are
no visual or morphological signs of interaction. (See Meyers11 for
detailed studies of these SN host galaxy properties.)

{\bf (c) Light curve:} SNe~Ia can be distinguished from most common
types of SNe~CC by some combination of light curve shape, color, and
absolute magnitude.  We compare the light curve of each candidate to
template light curves for SN~Ia and various SN~CC subtypes to test if
the candidate could be a SN~Ia or a SN~CC. For candidates lacking
both spectroscopic confirmation and an elliptical host galaxy, if
there is sufficient light curve data to rule out all SN~CC subtypes,
the candidate is considered a ``probable'' SN~Ia. If SN~Ia can be
ruled out, it is considered a ``probable'' SN~CC. If neither SN~Ia nor
SN~CC can be ruled out, the candidate is considered a ``plausible''
SN~Ia or SN~CC based on how typical the candidate's absolute magnitude
and/or color would be of each type. This approach can be viewed as a
qualitative version of the pseudo-Bayesian light curve typing
approaches of,
e.g., \citet{kuznetsova07a,kuznetsova08a,poznanski07b,poznanski07a}. SNe
classified as ``probable'' here would likely have a Bayesian posterior
probability approaching $1$, while ``plausible'' SNe would have an
intermediate probability (likely between 0.5 and 1.0). We consciously
avoid the full Bayesian typing approach because it can obscure large
uncertainties in the priors such as luminosity distributions, relative
rates, light curve shapes, and SN subtype fractions. Also, the
majority of our candidates have more available light curve information
than those of \citet{kuznetsova08a} and \citet{poznanski07a}, making a
calculation of precise classification uncertainty less necessary. In
general, classification uncertainty from light curve fitting is not a
concern for the cluster rate calculation as most cluster-member
candidates are securely typed using methods (a) and/or (b), above. It
is more of a concern for the volumetric field rate calculation based
on the non-cluster candidates (Barbary et~al., in preparation), though
the uncertainty in the field rate is still dominated by Poisson error.

For each candidate we fit template light curves for SN~Ia, Ibc, II-P,
II-L, and IIn. We use absolute magnitude and color as a discriminant
by limiting the allowed fit ranges according to the known
distributions for each subtype. For SN~Ia we start with the spectral
time series template of \citet{hsiao07a}, while for the core-collapse
types we use templates
of \citet{nugent02a}\footnote{See \url{http://supernova.lbl.gov/~nugent/nugent_templates.html}.}.
Each spectral time series is redshifted to the candidate redshift and
warped according to the desired color. Observer-frame template light
curves are then generated by synthetic photometry in the $i_{775}$ and
$z_{850}$ filters. The magnitude, color, date of maximum light, and
galaxy flux in $i_{775}$ and $z_{850}$ are allowed to vary to fit the
light curve data. For the SN~Ia template, the linear timescale or
``stretch'' \citep[e.g.,][]{perlmutter97a,guy05a} is also allowed to
vary within the range $0.6 < s < 1.3$.  We constrain the absolute
magnitude for each subtype to the range observed by \citet{li10a}; Our
allowed range fully encompasses their observed luminosity functions
(uncorrected for extinction) for a magnitude-limited survey for each
subtype. We correct from their assumed value of $H_0 =
73$~km~s$^{-1}$~Mpc$^{-1}$ to our assumed value of $H_0 =
70$~km~s$^{-1}$~Mpc$^{-1}$ and $K$-correct from $R$ to $B$ band. To
avoid placing too strict of an upper limit on SN~CC brightness, we use
the bluest maximum-light spectrum available when $K$-correcting (e.g.,
for SN Ibc we use a bluer spectrum than that of \citet{nugent02a}, as
bluer SNe Ibc have been observed). The resulting allowed $M_B$ range
for each subtype is shown in Table~\ref{table:snfit}. Note that the
range for Ibc does not include ultra-luminous SNe~Ic (such as those in
the luminosity functions of \citet{richardson02a}) as none were
discovered by \citet{li10a}. While such SNe can mimic a SN~Ia
photometrically, the \citet{li10a} results indicate that they are
intrinsically rare, and even \citet{richardson02a} show that they make
up at most $\sim$20\% of all SNe~Ibc. Still, we keep in mind that even
candidates compatible only with our SN~Ia template and incompatible
with SN~CC templates may in fact be ultra-luminous SNe~Ic, though the
probability is low. This is why any candidate typed based on light
curve alone has a confidence of at most ``probable,'' rather than
``secure.''  The allowed ranges of ``extinction,'' $E(B-V)$, are also
shown in Table~\ref{table:snfit}. For SN~Ia, $E(B-V)$ is the
difference in $B-V$ color from the \citet{hsiao07a} template. As the
observed distribution of SNe includes SNe
bluer than this template, SNe~Ia as blue as $E(B-V) = -0.2$ are
allowed. Given an $E(B-V)$, the spectral template is warped according
to the {\sc salt} color law \citep{guy05a}, with an effective $R_B =
2.28$ \citep{kowalski08a}. For SN~CC templates, extinction as low as
$E(B-V)=-0.1$ is allowed to reflect the possibility of SNe that are
intrinsically bluer than the \citet{nugent02a} templates. Templates
are then warped using a \citet{cardelli89a} law with $R_B =
4.1$. Extinctions are limited to $E(B-V) < 0.5$ (implying an
extinction of $A_B \sim 2$~magnitudes for SNe~CC).

\begin{deluxetable}{lcccc}
\tablewidth{0pt}
\tablecaption{\label{table:snfit} SN light curve template parameter ranges}
\tablehead{\colhead{SN type} & \colhead{Template} & \colhead{Observed $M_B$} & 
           \colhead{$E(B-V)$} & \colhead{$s$}}
\startdata
Ia   & Hsiao  & $-17.5$ -- $-20.1$ & $-0.2$ -- $0.6$ & $0.6$ -- $1.3$ \\
Ibc  & Nugent & $-15.5$ -- $-18.5$ & $-0.1$ -- $0.5$ & $1.0$\\
II-L & Nugent & $-16.0$ -- $-19.0$ & $-0.1$ -- $0.5$ & $1.0$\\
II-P & Nugent & $-15.5$ -- $-18.0$ & $-0.1$ -- $0.5$ & $1.0$\\
IIn  & Nugent & $-15.5$ -- $-19.1$ & $-0.1$ -- $0.5$ & $1.0$
\enddata
\end{deluxetable}

The light curve template with the largest $\chi^2$ $P$-value is
generally taken as the type. We also evaluate each fit by eye to check
that the best-fit template adequately describes the light
curve. Figure~\ref{fig:sn} shows the best-fit template for each
candidate.  For candidates typed on the basis of spectroscopic
confirmation or an elliptical host galaxy only the SN~Ia template is
shown. For candidates typed on the basis of the light curve alone, we
show both the best-fit SN~Ia and best-fit SN~CC templates for
comparison. The confidence in the best-fit template is either
``probable'' or ``plausible'' depending on how well other templates
fit: If the next-best fit has a $P$-value that is smaller than
$10^{-3} \times P_{\rm best}$, the best-fit template is considered the
only acceptable fit and the confidence is ``probable.'' If the
next-best fit has a $P$-value larger than $10^{-3} \times P_{\rm
best}$ the confidence is ``plausible.''  Finally, note that the
photometry used here is simple aperture photometry with fixed aperture
corrections. For SN~Ia cosmology we use color-dependent aperture
corrections, as described in Suzuki et al. (in preparation).

\begin{figure*}
\epsscale{1.175}
\plotone{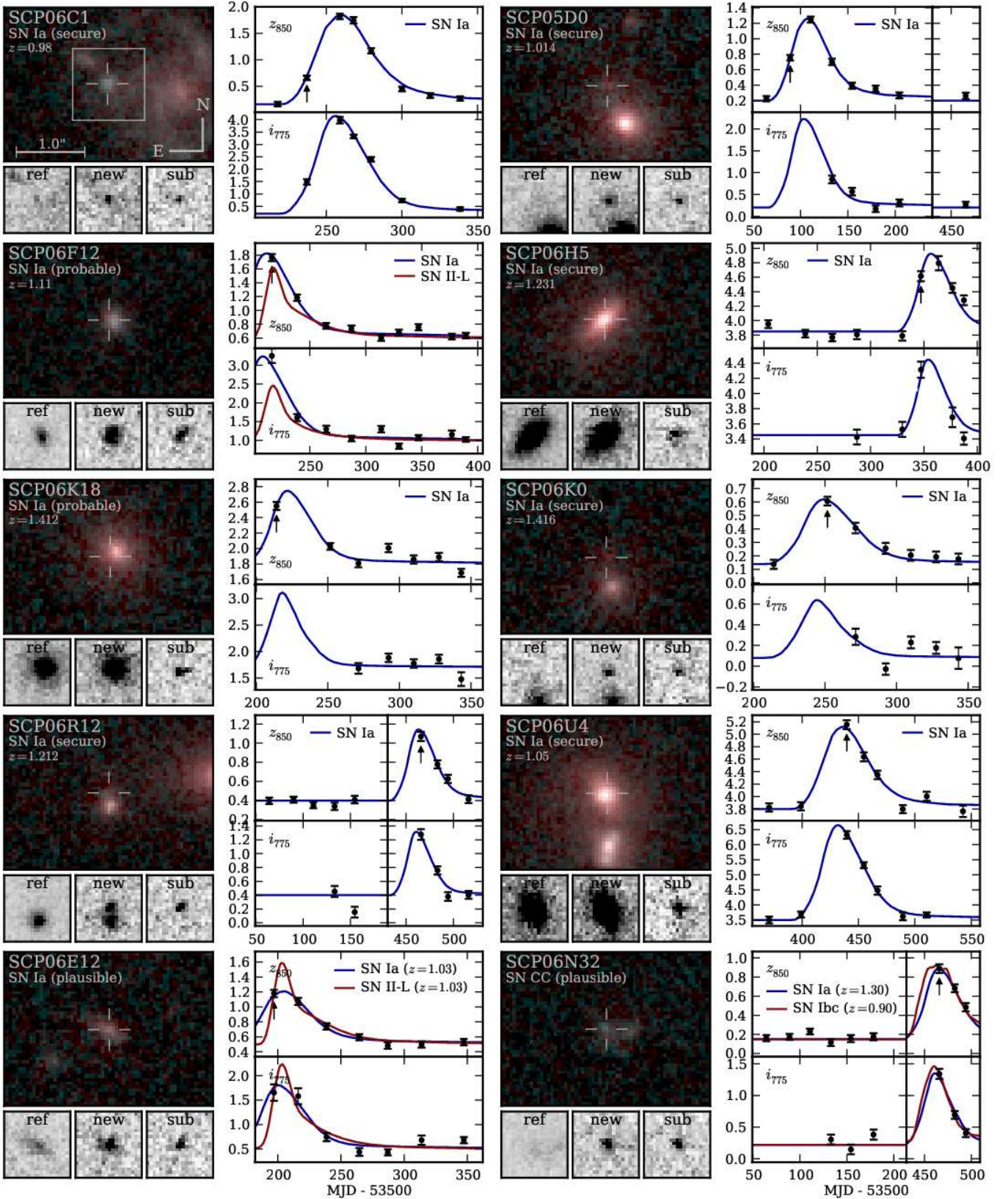}
\caption{Images and light curves of the 29 candidates classified as
supernovae. For each candidate, the upper left panel shows the 2-color
stacked image ($i_{775}$ and $z_{850}$) of the supernova host galaxy,
with the SN position indicated. The three smaller panels below the
stacked image show the reference, new, and subtracted images for the
discovery visit. The right panel shows the light curve at the SN
position (including host galaxy light) in the $z_{850}$ ({\it top})
and $i_{775}$ ({\it bottom}) bands. The y axes have units of counts
per second in a $3$~pixel radius aperture. The effective zeropoints
are 23.94 and 25.02 for $z_{850}$ and $i_{775}$, respectively. The
discovery visit is indicated with an arrow in the $z_{850}$ plot. The
best-fit SN~Ia template is shown in blue. For cases where the type is
SN~Ia based on spectroscopic confirmation or host galaxy environment,
only the best-fit SN~Ia template is shown, to demonstrate the
consistency of the light curve with the designation. For cases where
the type is based only on the light curve fit, the best-fit core
collapse SN template is shown in red.  Note that the photometry used
here is simple aperture photometry with fixed aperture
corrections. For SN~Ia cosmology we use color-dependent aperture
corrections, as described in Suzuki et al. (in preparation).
\label{fig:sn}}
\end{figure*}

\begin{figure*}
\figurenum{4}
\epsscale{1.175}
\plotone{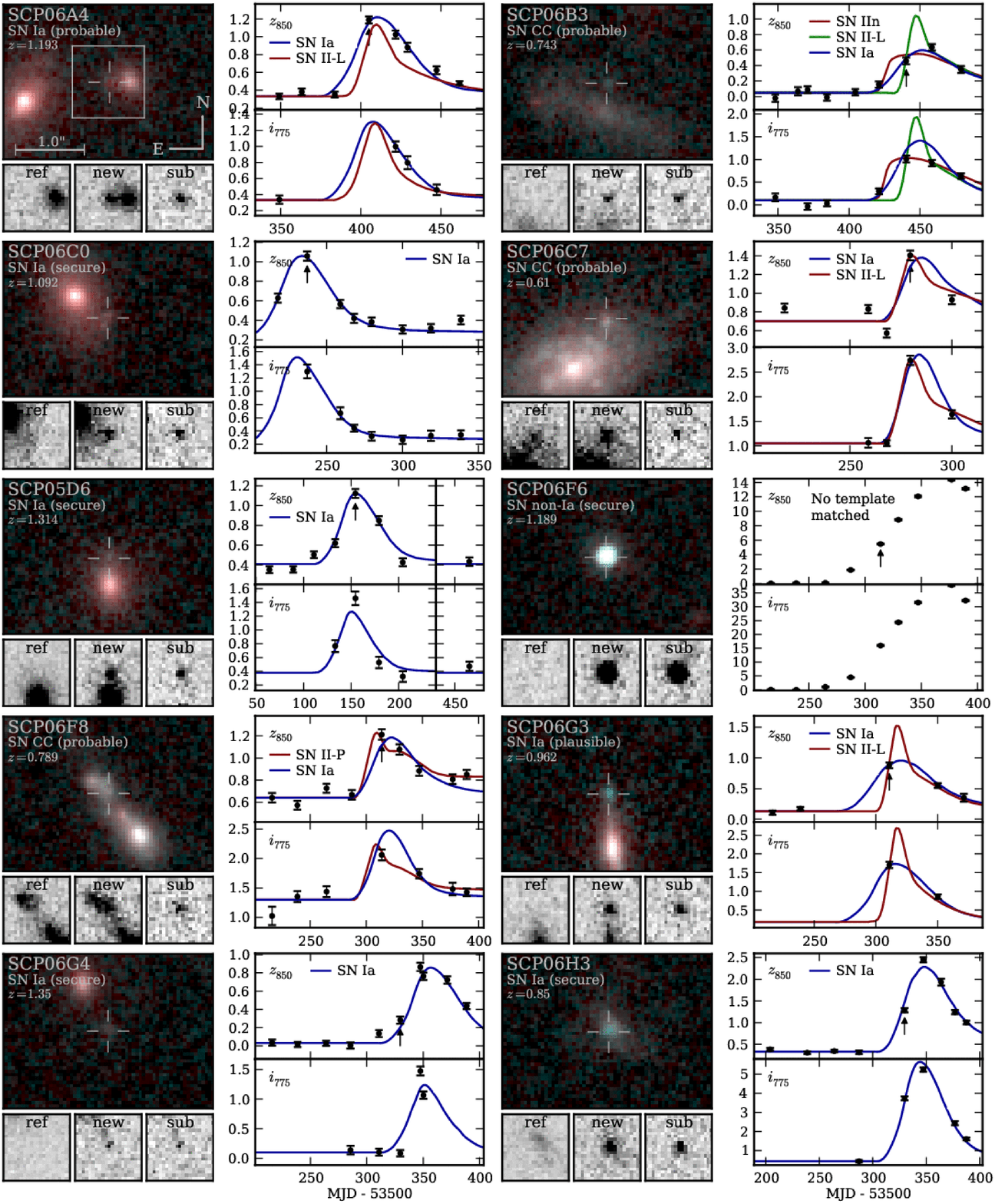}
\caption{\emph{Continued}}
\end{figure*}

\begin{figure*}
\figurenum{4}
\epsscale{1.175}
\plotone{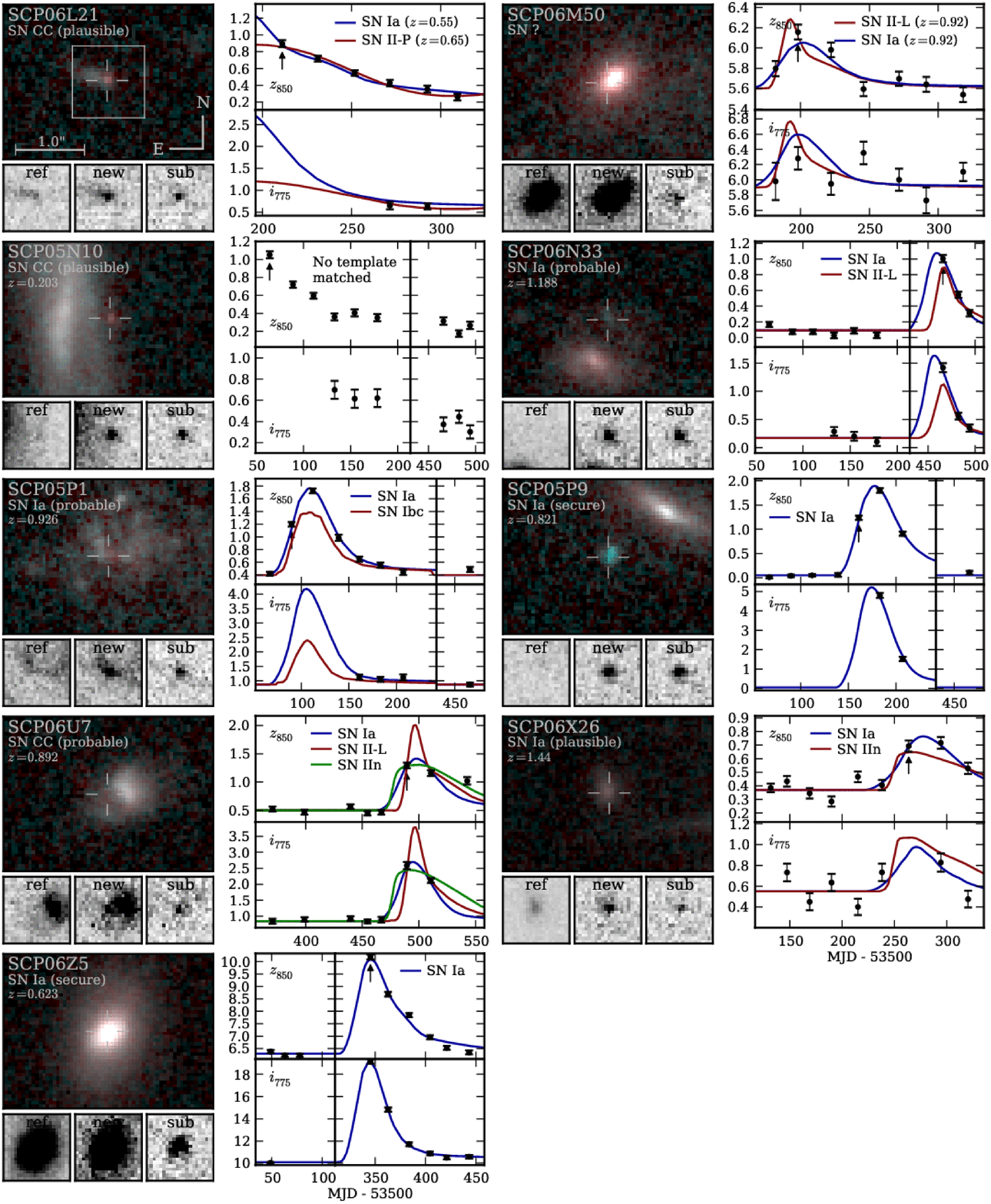}
\caption{\emph{Continued}}
\end{figure*}

\subsubsection{Comments on individual SN light curves} \label{sncomments}

Here we comment in greater detail on a selection of individual
candidates, particularly those with the greatest uncertainty in
typing. For each candidate, see the corresponding panel of
Figure~\ref{fig:sn} for an illustration of the candidate host galaxy and
light curve.

{\it SN SCP06E12}. We were unable to obtain a host galaxy redshift due
to the faintness of the host. The color of the host galaxy is
consistent with the cluster red sequence. The candidate light curve is
consistent with a SN~Ia at the cluster redshift of $z=1.03$, but is
also consistent with SN II-L at $z=1.03$. Different SN types provide
an acceptable fit over a fairly wide range of redshifts. As the SN~Ia
template provides a good fit with typical parameters, we classify the
candidate as a ``plausible'' SN~Ia.  However, there is considerable
uncertainty due to the uncertain redshift.

{\it SN SCP06N32} also lacks a host galaxy redshift. If the cluster
redshift of $z=1.03$ is assumed, the candidate light curve is best fit
by a SN~Ibc template. A SN~Ia template also yields an acceptable
fit, but requires an unusually red color of $E(B-V) \sim 0.6$. Given
the best-fit $s$ and $M_B$ values, the candidate would have an
unusually large Hubble diagram residual of approximately
$-0.8$~magnitudes. If the redshift is allowed to float, a SN~Ia
template with more typical parameters provides an acceptable fit at $z
= 1.3$. A SN~Ibc template still provides a better fit, with the best
fit redshift being $z \sim 0.9$. As SN~Ibc
provides a better fit in both cases, we classify this as a ``plausible''
SN~CC. However, there is considerable uncertainty in
both the type and cluster membership of this candidate. 

{\it SN SCP06A4}. We note that this candidate was observed
spectroscopically, as reported in Dawson09. While the spectrum was
consistent with a SN~Ia, there was not enough evidence to
conclusively assign a type. The host galaxy is morphologically and
photometrically consistent with an early-type galaxy, but there is
detected [OII], a possible indication of star formation. We therefore
rely on light curve typing for this candidate, assigning a confidence
of ``probable'' rather than ``secure.''
 
{\it SN SCP06G3} has only sparse light curve coverage. The best fit
template is a SN~Ia with $s=1.3$, $E(B-V)=0.3$ and $M_B=-18.5$, although
these parameters are poorly constrained. A large stretch and red
color would not be surprising given the spiral nature of the host
galaxy. It is also consistent with a II-L template, although the best
fit color is unusually blue: $E(B-V)=-0.1$. Given that SN~Ia yields
more ``typical'' fit parameters and that, at $z \sim 1$ a detected SN
is more likely to be Type~Ia than II, we classify this as a
``plausible'' Type Ia, with considerable uncertainty in the type.

{\it SN SCP06L21} lacks a spectroscopic redshift, but has a distinct
slowly-declining light curve that rules out a $z>0.6$ SN~Ia light
curve. Even the best-fit Ia template at $z=0.55$, shown
in Fig.~\ref{fig:sn}), is unusually dim ($M_B \approx -17.5$), making it
unlikely that the candidate is a lower-redshift SN~Ia. The light curve
is better fit by a SN~II-P template (with the best-fit redshift being
$z=0.65$). We therefore classify the candidate as a ``probable''
SN~CC.

{\it SN SCP06M50} is the most questionable ``SN'' candidate, having no
obvious $i_{775}$ counterpart to the increase seen in $z_{850}$. It
may in fact be an image artifact or AGN. However, it appears to be off
the core of the galaxy by $\sim$2~pixels (making AGN a less likely
explanation), and shows an increase in $z_{850}$ flux in two
consecutive visits, with no obvious cosmic rays or hot pixels (making
an image artifact less likely as well). The galaxy is likely to be a
cluster member: its color and magnitude put it on the cluster red
sequence, it is morphologically early-type, and it is only $19''$ from
the cluster center. Under the assumption that the candidate is a
supernova and at the cluster redshift of $z=0.92$, no template
provides a good fit due to the lack of an $i_{775}$ detection and the
constraints on $E(B-V)$. In particular, a SN~Ia template would
require $E(B-V)>0.6$. (The best-fit template shown in
Fig.~\ref{fig:sn} is with $E(B-V) = 0.6$.) If the redshift is allowed
to float, it is possible to obtain a good fit at higher redshift
($z \sim 1.3$), but still with $E(B-V) \gtrsim 0.4$, regardless of the
template type. Given the color and early-type morphology of the host
galaxy, it is unlikely to contain much dust. There is thus no
consistent picture of this candidate as a SN, and we do not assign a
type. However, note that the candidate is unlikely to be a cluster
SN~Ia.

{\it SN SCP05N10} is the lowest-redshift SN candidate in our sample at
$z=0.203$. Its light curve shape is inconsistent with a SN~Ia occurring
well before the first observation, and its luminosity is too low for
a SN~Ia with maximum only slightly before the first
observation. Therefore, we call this a ``probable'' SN CC. For all SN
types, the best fit requires maximum light to occur well before the
first observation, making all fits poorly constrained.

{\it SN SCP06X26} has a tentative redshift of $z=1.44$, derived from a
possible [OII] emission line in its host galaxy. Given this redshift,
a Ia template provides an acceptable fit, consistent with a typical
SN~Ia luminosity and color. However, we consider this a ``plausible,''
rather than ``probable, '' SN~Ia, given the uncertain redshift and low
signal-to-noise of the light curve data.

\subsection{Summary} \label{candsummary}
In the previous section we addressed the type of all 29 candidates
thought to be SNe. However only the cluster-member SNe~Ia are of
interest for the remainder of this paper. There are six ``secure''
cluster-member SNe~Ia, and two ``probable'' SNe~Ia, for a total of
eight. In addition, SCP06E12 is a ``plausible'' SN~Ia and may be a
cluster member. Two other candidates, SCP06N32 and SCP06M50, cannot be
definitively ruled out as cluster-member SNe~Ia, but are quite
unlikely for reasons outlined above. We take eight cluster SNe~Ia as
the most likely total. It is unlikely that {\it both} of the
``probable'' SNe~Ia are in fact SNe~CC. We therefore assign a
classification error of $^{+0.0}_{-0.5}$ for each of these, resulting
in a lower limit of seven cluster-member SNe~Ia. There is a good
chance that SCP06E12 is a cluster-member SN~Ia, while there is only a
small chance that SCP06N32 and SCP06M50 are either cluster SNe~Ia. For
these three candidates together, we assign a classification error of
$^{+1}_{-0}$, for an upper limit of nine. Thus, $8 \pm 1$ is the total
number of observed cluster SNe~Ia.

\section{Effective Visibility Time} \label{ct}

With a systematically selected SN~Ia sample now in hand, the cluster
SN~Ia rate is given by
\begin{equation}
\label{eq:rate}
\mathcal{R} = \frac{N_{\rm SN~Ia}}{\sum_j T_j L_j} ,
\end{equation}
where $N_{\rm SN~Ia}$ is the total number of SNe~Ia observed in
clusters in the survey, and the denominator is the total effective
time-luminosity for which the survey is sensitive to SNe~Ia in
clusters. $L_j$ is the luminosity of cluster $j$ visible to the survey
in a given band. $T_j$ is the ``effective visibility time'' (also
known as the ``control time'') for cluster $j$. This is the effective
time for which the survey is sensitive to detecting a SN~Ia,
calculated by integrating the probability of detecting a SN~Ia as a
function of time over the span of the survey. It depends on the
redshift of the SN~Ia to be detected and the dates and depths of the
survey observations. As each cluster has a different redshift and
different observations, the control time is determined separately for
each cluster.  To calculate a rate per stellar mass, $L_j$ is replaced
by $M_j$.

Equation~(\ref{eq:rate}) is for the case where the entire observed
area for each cluster is observed uniformly, yielding a control time
$T$ that applies to the entire area.  In practice, different areas of
each cluster may have different observation dates and/or depths,
resulting in a control time that varies with position. This is
particularly true for this survey, due to the rotation of the observed
field between visits and the gap between ACS chips. Therefore, we
calculate the control time as a function of position in each observed
field, $T_j(x,y)$. As the cluster luminosity is also a function of
position, we weight the control time at each position by the
luminosity at that position. In other words, we make the substitution
\begin{equation} 
\label{eq:ratedenom}
T_j L_j \Rightarrow \int_{x,y} T_j (x,y) L_j (x,y). 
\end{equation}

The effective visibility time $T$ at a position $(x,y)$ on the 
sky is given by
\begin{equation}
T(x,y) = \int_{t=-\infty}^{t=\infty} \eta^\ast (x,y,t) \epsilon (x,y,t) dt.
\end{equation}
The integrand here is simply the probability for the survey and our
selection method to detect (and keep) a SN~Ia at the cluster redshift
that explodes at time $t$, and position $(x,y)$. This probability is
split into the probability $\eta^\ast$ of detecting the supernova and
the probability $\epsilon$ that the supernova passes all ``light
curve'' cuts. As each SN has multiple chances for detection, the total
probability of detection $\eta^\ast$ is a combination of the
probabilities of detection in each observation. For example, if we
have two search visits at position $(x,y)$, $\eta^\ast(t)$ is given by
\begin{equation}
\eta^\ast (t) = \eta_1 (t)+ ( 1-\eta_1(t) ) \eta_2 (t),
\end{equation}
where $\eta_i (t)$ is the probability of detecting a SN~Ia 
exploding at time $t$ in visit $i$. In other words, the total 
probability of finding the SN~Ia exploding at time $t$ is the probability 
of finding it in visit 1 plus the probability that it was \emph{not} found in
visit 1 times the probability of finding it in visit 2. This can be 
generalized to many search visits: The contribution of each additional visit 
to the total probability is the probability of not finding the SN in any 
previous visit times the probability of finding the SN in that visit.

In practice, we calculate $T(x,y)$ in two steps: First, we determine
the probability $\eta$ of detecting a new point source in a single
image as a function of the point source magnitude. This is discussed
in \S\ref{eff}. Second, for each $(x,y)$ position in the observed area
we simulate a variety of SN~Ia light curves at the cluster redshift
occurring at various times during the survey. By considering the dates
of the observations made during the survey at that specific position,
we calculate the brightness and significance each simulated SN~Ia would
have in each $z_{850}$ and $i_{775}$ image. We then use our
calculation of $\eta$ as a function of magnitude to convert the
observed brightness into a probability of detecting the simulated SN
in each observation. The light curve simulation is discussed in
\S\ref{lightcurves}. The calculation of cluster luminosities,
$L_j(x,y)$, is discussed in \S\ref{lum}.

\subsection{Detection Efficiency Versus Magnitude} \label{eff}

Here we calculate the probability of detecting a new point source as a
function of magnitude in a single subtraction. We use a Monte Carlo
simulation in which artificial point sources of various magnitudes are
added to each of the individual exposure images from the survey,
before they are combined using {\sc MultiDrizzle}. Starting from the
individual exposures allows us to test both the efficiency of the {\sc
MultiDrizzle} process and our cosmic ray rejection (which uses the
flux observed in the individual exposures). The point sources are
placed on galaxies in positions that follow the distribution of light
in each galaxy. Poisson noise is added to each pixel in the point
source. The altered images are then run through the full image
reduction and SN detection pipeline used in the search, and flagged
candidates are compared to the input point sources.

\begin{figure*}
\epsscale{1.175}
\plotone{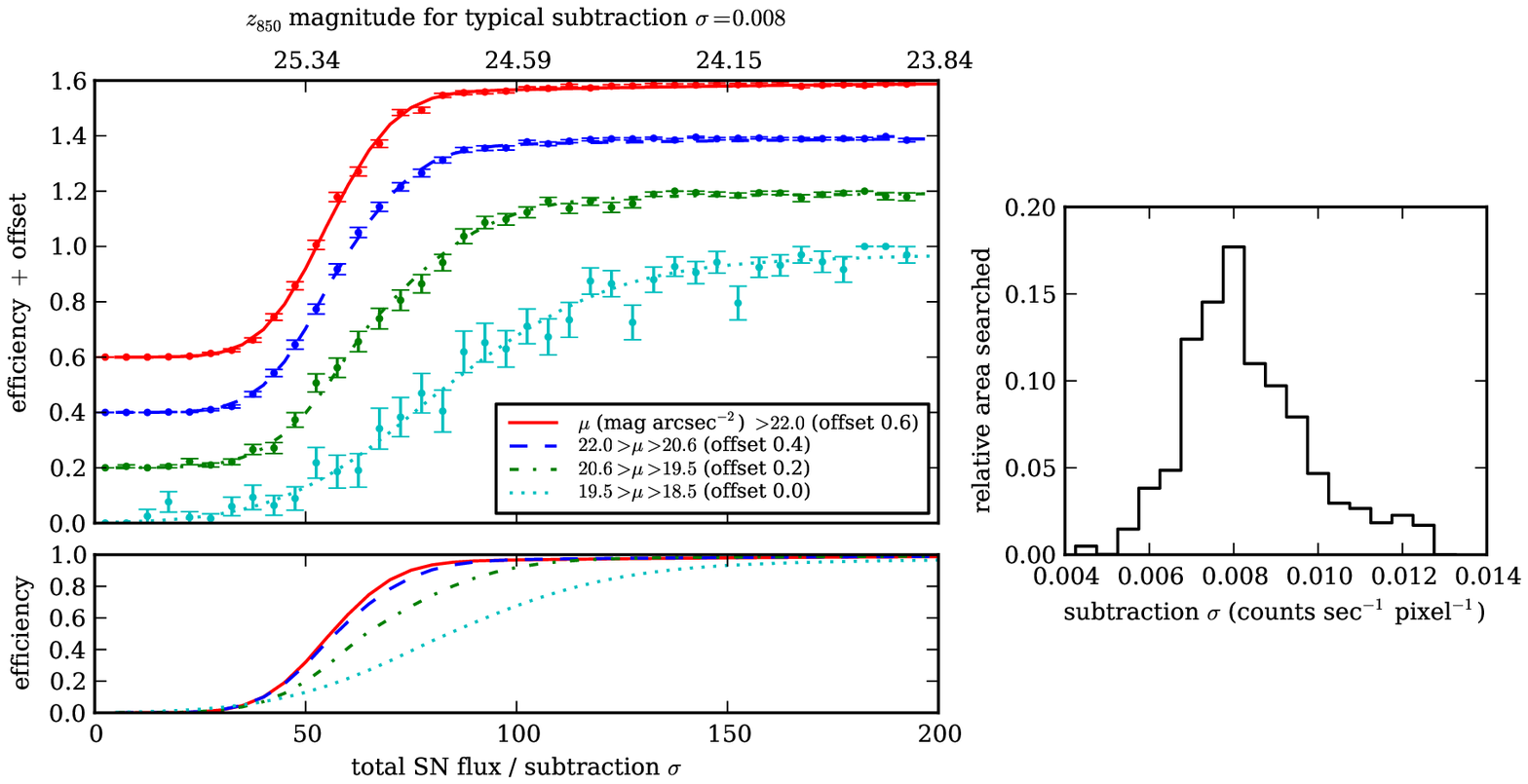}
\caption{Point source detection efficiency in a single subtraction, 
as a function of the ratio of total point source flux to subtraction
noise $\sigma$ (counts~sec$^{-1}$~pixel$^{-1}$). The artificial point
sources are split into four bins depending on the underlying galaxy
surface brightness $\mu$ (mag~arcsec$^{-2}$) at the point source
position. The efficiency curve is calculated separately for each
bin. In the upper left panel, the four bins are shown, offset for
clarity. In the lower left panel, the fitted curves are reproduced
without offset for comparison. Approximately 72,000 artificial point
sources were used in total. The right panel shows the distribution of
the noise level in the subtractions. The noise level differs by a
factor of about two from the deepest to shallowest subtractions
searched. 
\label{fig:eff}}
\end{figure*}

We parameterize the detection efficiency by the ratio of point source
flux to sky noise. This is a good choice because, in most cases, the
detection efficiency will depend only on the contrast between the
point source and the sky noise. However, there is an additional
dependence on the surface brightness at the location of the point
source: point sources near the core of galaxies will have a lower
detection efficiency due to additional Poisson noise from the
galaxy. For $0.6<z<1.5$ galaxies, we estimate that only $\sim$$10\%$
of SNe will fall on regions where galaxy Poisson noise is greater than
the sky noise (assuming SNe follow the galaxy light
distribution). Still, we take this effect into account by splitting
our sample of artificial point sources into four bins in underlying
surface brightness. The detection efficiency is calculated separately
in each bin (Fig.~\ref{fig:eff}, top left panel).  The first two bins,
$\mu > 22.0$ and $22.0 > \mu > 20.6$ mag~arcsec$^{-2}$, correspond to
lower surface brightnesses where sky noise is dominant. As expected,
their efficiency curves are very similar. In the third and fourth
bins, corresponding to higher surface brightness, the Poisson noise
from the galaxy dominates the sky noise, and the efficiency drops as a
result.

For reference, the distribution of sky noise in the subtractions is
shown in Figure~\ref{fig:eff} (right panel). Nearly all the searched
area has a sky noise level between 0.006 and
0.012~counts~sec$^{-1}$~pixel$^{-1}$. For a typical value of 0.008, we
show the corresponding point source $z_{850}$ magnitude on the top
axis of the left panel.

We find that the efficiency curve in each bin is well-described by
the function
\begin{equation}
\eta(x) = \left\{ \begin{array}{ll}
\frac{1}{2} (1+ae^{-bx}) [\mathrm{erf}((x-c)/d_1)+1], & x<c     \\
\frac{1}{2} (1+ae^{-bx}) [\mathrm{erf}((x-c)/d_2)+1], & x \ge c
\end{array} \right. ,
\end{equation}
where $x$ is the ratio of point source flux to sky noise, and $a$,
$b$, $c$, $d_1$ and $d_2$ are free parameters. An error function is
the curve one would expect with a constant cut and Gaussian noise, but
we find that two different scales ($d_1$ and $d_2$) in the error
function, as well as an additional exponential term, are necessary to
describe the slow rise to $\eta =1$ at large $x$. This slow rise is
due to rarer occurrences, such as cosmic rays coinciding with new point
sources. The fitted functions for the four bins are plotted in the top
left of Figure~\ref{fig:eff} and reproduced in the bottom left of the
figure for comparison. We use these fitted functions to calculate the
effective visibility time in the following section. 

\subsection{Simulated Lightcurves} \label{lightcurves}

\begin{figure}
\epsscale{1.175}
\plotone{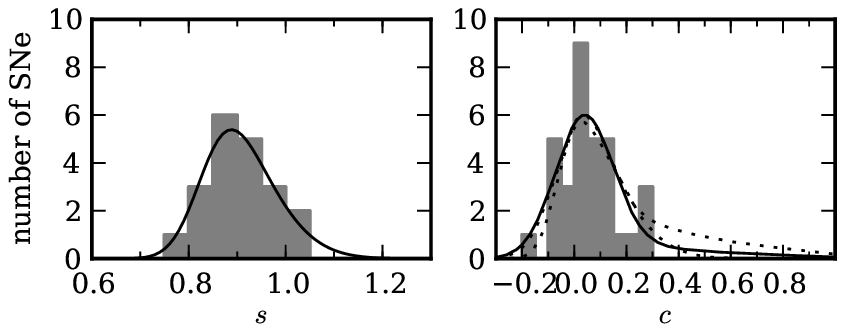}
\caption{{\it Left panel:} stretch distribution used for simulated SNe 
(\emph{solid line}) and the stretch distribution of first-year SNLS
$z<0.75$ SNe in passive hosts \citep{sullivan06a} (\emph{grey
histogram}). Note that the distribution is not changed significantly
by cutting the sample at $z<0.6$. Therefore we do not expect the
sample to be significantly Malmquist biased. {\it Right panel:} color
distribution of the first-year SNLS $z<0.6$ SNe \citep{astier06a}
(\emph{grey histogram}) and the color distribution used for simulated
SNe (\emph{solid line}). The \emph{dotted lines} show alternative
color distributions used to assess the possible systematic error due
to varying amounts of SNe being affected by dust.\label{fig:dists}}
\end{figure}

We simulate SN~Ia light curves with a distribution of shapes, colors
and absolute magnitudes.  We use the (original) {\sc
salt} \citep{guy05a} prescription in which the diversity of SN~Ia
light curves is characterized as a two-parameter family with an
additional intrinsic dispersion in luminosity.  The two parameters are
the linear timescale of the light curve (``stretch'', $s$) and the
$B-V$ color excess, $c$.  For each simulated SN, $s$ and $c$ are
randomly drawn from the distributions shown in Figure~\ref{fig:dists}
(solid lines). The stretch distribution is based on the
observed distribution in passive hosts
(Fig.~\ref{fig:dists}, left panel, grey histogram) in the
first-year Supernova Legacy Survey (SNLS) sample \citep{sullivan06a}.
Similarly, the color distribution is based on the observed color
distribution (Fig.~\ref{fig:dists}, right panel, grey
histogram) in the first-year SNLS sample \citep{astier06a}.  The
absolute magnitude of each simulated SN is set to
\begin{equation}
M_B = -19.31 - \alpha (s-1) + \beta c + I
\end{equation}
where $-19.31$ is the magnitude of an $s=1$, $c=0$ SN~Ia in our
assumed cosmology \citep{astier06a}, $\alpha = 1.24$, $\beta =
2.28$ \citep{kowalski08a}, and $I$ is an added ``intrinsic
dispersion'', randomly drawn from a Gaussian distribution centered at
zero with $\sigma = 0.15$~mag.

We have chosen distributions that represent as accurately as possible
the full distribution of SNe~Ia occurring in reality. However, note
that the control time is not actually very sensitive to the assumed
distributions. This is because, for the majority of cluster redshifts
in the survey, the detection efficiency is close to 100\% during the
time of the survey. Supernovae would thus have to be significantly
less luminous in order to change the detection efficiency
significantly. In the following section \S\ref{ctsys} we quantify the
effect on the control time arising from varying the assumed SN~Ia
properties and show that they are sub-dominant compared to the Poisson
error in the number of SNe observed. All sources of systematic errors
are also summarized in \S\ref{resultssys}.

To generate the simulated light curves in the observed bands, we use
the \citet{hsiao07a} SN~Ia spectral time series template. For each
simulated SN, the spectral time series is warped to match the selected
color $c$ and redshifted to the cluster restframe. Light curves are
generated in the observed $i_{775}$ and $z_{850}$ filters using
synthetic photometry, and the time axis is scaled according to the
chosen value of $s$.

\begin{figure*}
\epsscale{1.15}
\plottwo{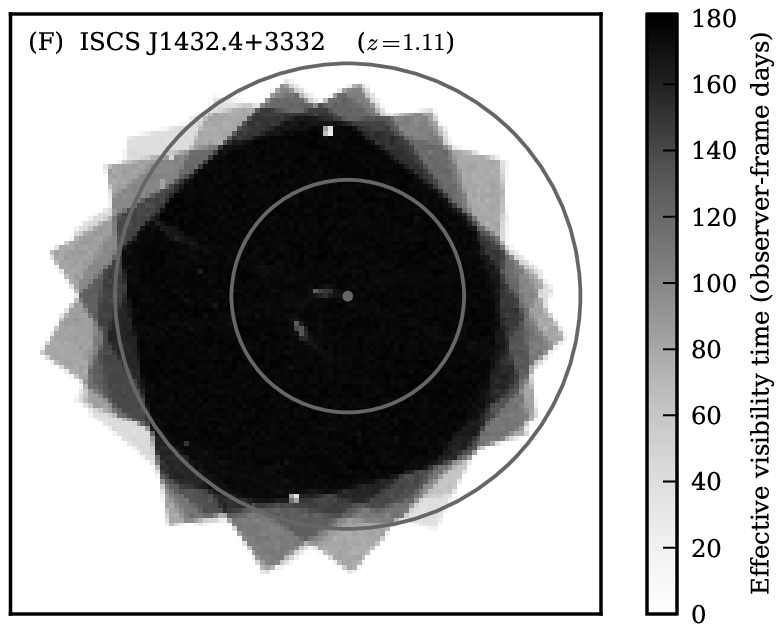}{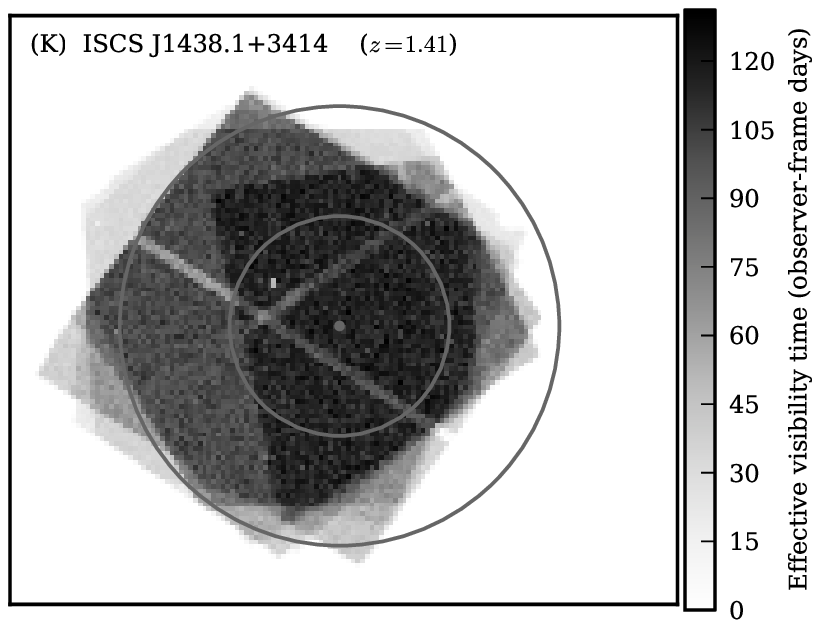}
\caption{Example maps of effective visibility time for clusters 
ISCS J1432.4+3332 (F) and ISCS J1438.1+3414 (K).  The dot denotes the
cluster center and the inner and outer circles represent 0.5~Mpc and
1.0~Mpc radius, respectively. The ``noise'' in these maps is due to
the finite number (100) of SNe simulated at each position.  At lower
redshift nearly all simulated SNe are recovered at each position,
whereas at higher redshift a sizable fraction of simulated SNe are
missed, resulting in a higher ``noise'' level.
\label{fig:ctmaps}}
\end{figure*}

For each cluster, we calculate $T(x,y)$ in bins of 50 $\times$
50~pixels ($2''.5\ \times\ 2''.5$). In each bin, we simulate 100 SN
light curves at random positions within the bin.  For each simulated
SN light curve, we shift the light curve in time across the entire
range of observations, starting with maximum light occurring 50~days
before the first observation and ending with maximum light occurring
50~days after the last observation. For each step in time we get the
$z_{850}$ and $i_{775}$ magnitude of the SN at every date of
observation. From the sky noise maps, we know the noise at the
position of the simulated SN in every image. Using the curves in
Figure~\ref{fig:eff}, we convert the SN flux-to-noise ratio to the
probability of the SN being detected in each $z_{850}$ exposure. (Each
simulated SN is also assigned a host galaxy surface brightness chosen
from a distribution, in addition to the randomly selected $s$, $c$ and
$I$ parameters; we use the Fig.~\ref{fig:eff} curve that corresponds
to this surface brightness.) At the same time, we calculate the
probability that the SN passes our light curve cuts (using both
$z_{850}$ and $i_{775}$ simulated magnitudes). Multiplying these two
probabilities gives the total probability of the simulated SN being
included in the sample if it peaks at the given date.  Integrating the
probability over time (the entire range of dates) gives the control
time for each simulated SN. We take the average control time of the
100 SNe as the value for the given bin. The resulting control time
map, $T(x,y)$, therefore has a resolution of $2''.5\ \times\
2''.5$. $T(x,y)$ is shown for two example clusters in
Figure~\ref{fig:ctmaps}.

\subsection{Effect of Varying SN Properties} \label{ctsys}

If the real distributions of SN~Ia properties differs significantly
from those assumed in our simulation, the $T(x,y)$ maps we have
derived could misrepresent the true efficiency of the survey. Above we
argued that the effect is likely to be small because the detection
efficiency is close to 100\% for most of the survey. Here we quantify
the size of the possible effect on the control time by varying the
assumed distributions.

To first order, changing the assumed distributions of $s$ or $c$ or
changing the assumed spectral time series will affect the detection
efficiency by increasing or decreasing the luminosity of the simulated
SN. To jointly capture these effects, we shift the absolute magnitude
of the simulated SNe~Ia by $^{+0.2}_{-0.2}$~mag and recalculate the
control times. To first order, this is equivalent to shifting the $s$
distribution by $\Delta s = 0.2/\alpha \sim 0.16$ or shifting the $c$
distribution by $\Delta c = 0.2/\beta \sim 0.09$. A $-0.2$~mag shift
in absolute magnitude increases the control time, decreasing the
inferred SN~Ia rate by $6\%$. A $+0.2$~mag shift decreases the control
time, increasing the SN~Ia rate by $8\%$. These effects are
sub-dominant compared to the Poisson error of $\gtrsim 30\%$ in the
number of SNe observed. (Sources of error are summarized
in \S\ref{resultssys} and Table~\ref{table:clratesys}.)

For the color distribution, in addition to a simple shift, we also
quantify the effect of including a smaller or larger fraction of SNe
significantly reddened by dust. In fact, we have good reasons to
believe that most cluster SNe~Ia will be in dust-free environments. A
large fraction of the stellar mass in the clusters ($\sim 80\%$) is
contained in red-sequence galaxies expected to have little or no
dust. Our spectroscopic and photometric analysis (Meyers11) of the
red-sequence galaxies confirms this expectation. Therefore, for our
default $c$ distribution (Fig.~\ref{fig:dists}, right panel, solid
line), we assumed that $20\%$ of SNe (those occurring in galaxies not
on the red sequence) could be affected by dust, and that the
extinction of these SNe would be distributed according to
$P(A_V) \propto \exp(-A_V/0.33)$ [the inferred underlying $A_V$
distribution of the SDSS-II sample \citep{kessler09a}]. All SNe are
assumed to have an intrinsic dispersion in color to match the observed
SNLS distribution at $c<0.3$. It might be the case that even fewer SNe
are affected by dust, or (unlikely) more SNe are affected by dust. As
extreme examples, we tested two alternative distributions (dotted
lines in Fig.~\ref{fig:dists}). In the first, we assumed that the SNLS
sample was complete and characterized the full $c$ distribution, with
a negligible number of $c>0.4$ SNe. This increases the control time by
only $2\%$. In the second, we increase the fraction of dust-affected
SNe from $20\%$ to $50\%$.  Even though this alternative distribution
includes an additional $\sim$$30\%$ more reddened SNe (unlikely to be
true in reality), the average control time is only lower by $9\%$
(increasing the rate by $10\%$). We use these values as the systematic
error in the assumed dust distribution.

\section{Cluster Luminosities and Masses} \label{lum}

In this section, we calculate the total luminosity of each cluster and
use the luminosity to infer a stellar mass. Only a small subset of
galaxies in each field have known redshifts, making it impossible to
cleanly separate cluster galaxies from field galaxies.  Therefore, we
use a ``background subtraction'' method to estimate cluster
luminosities statistically: we sum the luminosity of all detected
galaxies in the field and subtract the average ``background
luminosity'' in a non-cluster field.  This approach follows that
of \citet{sharon07a}.  For the blank field, we use the
GOODS\footnote{Based on observations made with the NASA/ESA {\it
Hubble Space Telescope}. The observations are associated with programs
GO-9425, GO-9583 and GO-10189} fields \citep{giavalisco04a} as they
have similarly deep or deeper observations in both ACS $i_{775}$ and
$z_{850}$.  In \S\ref{lumphotometry} we describe the galaxy detection
and photometry method.  Simply summing the photometry from the
detected galaxies would include most of the total cluster
light. However, for an unbiased estimate of the total light, several
small corrections are necessary: We account for light in the outskirts
of each galaxy (\S\ref{lumcorrection}), and light from faint galaxies
below the detection threshold (\S\ref{lumfaintgals}). These
corrections are on the order of 20\% and 5\%
respectively. In \S\ref{lumkcorrections} we convert the observed
$z_{850}$ flux to a rest-frame $B$-band flux. In \S\ref{lumprofile} we
sum the light and subtract background light. In \S\ref{lumsubsets} we
repeat this calculation limiting ourselves to red-sequence and
red-sequence early-type subsets of galaxies. Finally,
in \S\ref{lummass} we estimate cluster stellar masses based on the
cluster luminosities and stellar mass-to-light ratios.

\subsection{Galaxy Selection and Photometry} \label{lumphotometry}

We use the stacked $i_{775}$ and $z_{850}$ band images of each
cluster, which have total exposure times in the range 1060 --
4450~seconds and 5440 -- 16,935~seconds, respectively. Galaxy catalogs
are created using the method described in detail by Meyers11: We run {\sc
SExtractor} \citep{bertin96a} in dual-image mode using the $z_{850}$
image for detection, and use a two-pass Cold/Hot method \citep{rix04a}
to optimally de-blend galaxies. We remove stars from the catalog based
on the CLASS\_STAR and FLUX\_RADIUS parameters from the $z_{850}$
image.

It is notoriously difficult to determine accurate total fluxes for
extended sources. However, as we are only concerned with the summed
flux of many galaxies, it is not important that the estimate be
accurate for each individual galaxy, only that the estimate is
unbiased in the aggregate. We use the {\sc SExtractor} MAG\_AUTO photometry
(which gives the total flux within a flexible elliptical aperture) and
apply a correction determined using the Monte Carlo simulation described
below. In order to make the aperture correction as small as possible,
we use a relatively large ``Kron factor'' of 5.0, meaning that the
MAG\_AUTO aperture is scaled to 5.0 times the Kron radius of the
galaxy. MAG\_AUTO is only used to determine $z_{850}$ magnitudes;
$i_{775}-z_{850}$ colors are determined using PSF matching and a
smaller aperture, as described in Meyers11.

\subsection{Galaxy Detection Completeness and Magnitude Bias} 
\label{lumcorrection}

To count all the flux in all cluster galaxies, we must make two
corrections: (1) add the galaxy light outside of the MAG\_AUTO
aperture, and (2) add the luminosity of all cluster galaxies below the
detection threshold of our galaxy catalog. We use a Monte Carlo
simulation of galaxies placed on our real survey data to determine
both the detection efficiency as a function of galaxy magnitude, and
the fraction of galaxy light inside the MAG\_AUTO aperture. Each
simulated galaxy has a \citet{sersic68a} profile, with the S{\'e}rsic
index $n$ simply selected from a flat distribution ranging from $n =
0.7$ to $n = 4.5$, and the minor to major axis ratio $q$ selected from
a flat distribution ranging from $q = 0.3$ to $q = 1$.  The
distribution of galaxy angular sizes will also affect the results. For
guidance on the size of the galaxies of concern (namely, those at
$z \gtrsim 0.9$) we turned to the subsample of the 672 galaxies having
spectroscopic redshifts $0.85<z<1.6$. These 672 galaxies were all fit
with {\sc galfit} \citep{peng02a}, which fits a value for $r_e$. Based
on the distribution of $r_e$ as a function of magnitude for these
galaxies, we chose $r_e$ for each simulated galaxy (based on its
magnitude).  A total of 15000 and 12000 simulated galaxies were placed
on cluster and GOODS fields respectively.

\begin{figure}
\epsscale{1.175}
\plotone{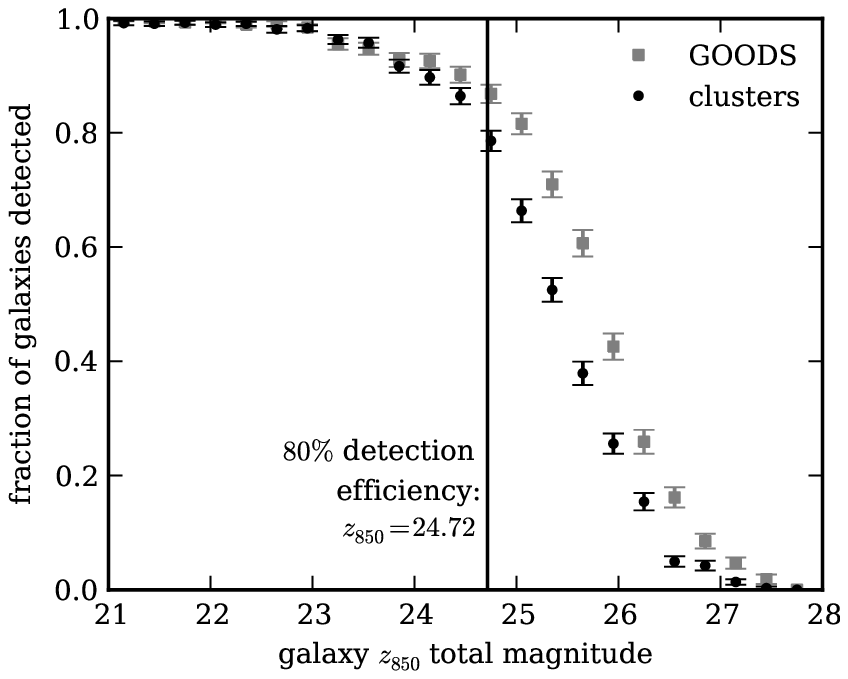}
\caption{Percentage of simulated galaxies recovered by {\sc SExtractor} as a 
  function of total galaxy $z_{850}$ magnitude for simulated galaxies
  placed on cluster fields ({\it black circles}) and GOODS fields
  ({\it grey squares}). The detection efficiency drops to 80\% at
  $z_{850} = 24.72$ for cluster fields (\emph{vertical line}). We discard all
  galaxies dimmer than this value.\label{fig:fakegals_eff}}
\end{figure}

The detection efficiency as a function of galaxy magnitude is shown in
Figure~\ref{fig:fakegals_eff}. For the average of all cluster fields,
the detection efficiency drops to 80\% at $z_{850} = 24.72$. We use
this magnitude as a cutoff in our selection, discarding all galaxies
dimmer than this magnitude.  We later correct total cluster
luminosities for the uncounted light from these galaxies by using an
assumed cluster luminosity function.  In reality, the detection
efficiency varies slightly from field to field (and even within a
field) due to exposure time variations. However, to first order, the
variation is accounted for by using the average efficiency in all
fields. In addition, the total luminosity of $z_{850} > 24.72$ cluster
galaxies is expected to be small (as we show below), so slight changes
in the cutoff will have a negligible effect on the total luminosity.

\begin{figure}
\epsscale{1.175}
\plotone{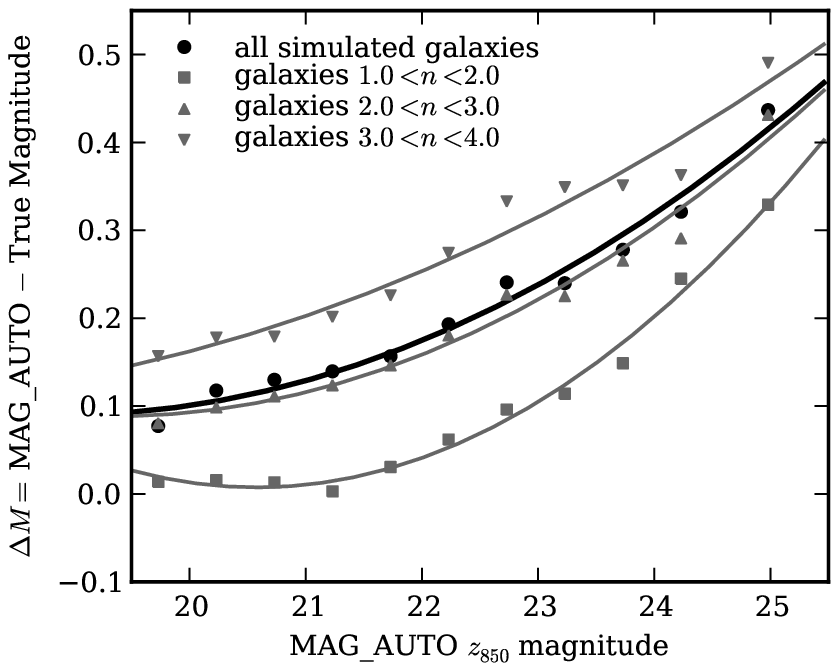}
\caption{Galaxy MAG\_AUTO aperture correction as a function of galaxy 
magnitude. {\it Black circles:} Average correction for the full
distribution of galaxies simulated, including all S{\'e}rsic indices
$n$. The {\it black line} is a fit to these points and is the relation
we use. Note that it is not extrapolated beyond the range shown. To
illustrate the effect of $n$ on the aperture correction, we plot the
aperture correction for subsets of galaxies with different S{\'e}rsic
indices ({\it Grey squares and triangles}). Galaxies with larger
S{\'e}rsic indices have a larger aperture correction.
\label{fig:fakegals_apcorr}}
\end{figure}

For each simulated galaxy, we determine the difference ($\Delta M$)
between the MAG\_AUTO magnitude and the true total magnitude. Binning
the simulated galaxies by their MAG\_AUTO magnitude, we derive a
relation between $\Delta M$ and the galaxy brightness
(Fig.~\ref{fig:fakegals_apcorr}, black circles). $\Delta M$
generally increases with galaxy magnitude because the outskirts of
dimmer galaxies are increasingly buried in noise, causing {\sc
SExtractor} to underestimate the true extent of the galaxy, and
thereby underestimate the Kron radius, resulting in a smaller
MAG\_AUTO aperture. We find that the relation is well-fit by a
second-order polynomial (Fig.~\ref{fig:fakegals_apcorr}, thick black
line), given by
\begin{eqnarray}
\Delta M & = & 0.238 + 0.081(M_{MAG\_AUTO}-23) + \nonumber \\
         &   & + 0.009(M_{MAG\_AUTO} -23)^2.
\end{eqnarray}
We use this to correct the magnitude of each detected galaxy. Note
that the correction is not extrapolated beyond the fitted range shown.

Because we cannot reliably determine $r_e$ or the S{\'e}rsic index $n$
for each galaxy, we rely on the simulated distribution of $r_e$ and
$n$ to accurately represent the true distributions. (The black circles
in Fig.~\ref{fig:fakegals_apcorr} include all simulated galaxies.)  We
have based our distribution of $r_e$ on actual galaxies, but $n$ is
less well-known. To estimate the effect of varying the $n$
distribution, we show $\Delta M$ for subsets of the simulated
galaxies, divided by S{\'e}rsic index (Fig.~\ref{fig:fakegals_apcorr},
grey points and lines). $\Delta M$ increases with S{\'e}rsic index,
because a larger S{\'e}rsic index implies a larger fraction of light
in the outskirts of the galaxy, under the detection threshold. This
leads to a smaller estimate of the Kron radius, and a smaller
MAG\_AUTO aperture.  If, instead of the flat $1<n<4$ distribution
used, all galaxies had $1<n<2$, the aperture correction would be lower
by approximately $0.10$~magnitudes. If instead all galaxies had
$3<n<4$, the correction would be higher by approximately
$0.07$~magnitudes. We use $0.07$~mag as the systematic uncertainty in
the aperture correction. (All systematic uncertainties are summarized
in \S\ref{resultssys} and Table~\ref{table:clratesys}.)

\subsection{$K$-Corrections} \label{lumkcorrections}

We use a $K$-correction based on the BC03 stellar population spectral
models to convert the observed $z_{850}$ magnitude to a rest-frame $B$
magnitude for each cluster. Rather than using a single $K$-correction
for all the light in each cluster, we apply a $K$-correction to each
galaxy magnitude based on its $i_{775}-z_{850}$ color.  For each
cluster's redshift, we determine the relation between $K$-correction
($M_B$ (rest) $- z_{850}$) and $i_{775}-z_{850}$ color, using BC03
spectra with initial metallicities in the range $0.004 < Z < 0.05$ and
ages in the range $1 \times 10^8 - 5 \times 10^9$~yr. For most cluster
redshifts in our sample, all of the spectra over this wide range fall
along the same line in $K$-correction versus color, meaning that the
color determines the $K$-correction, regardless of the metallicity or
age assumed. The dispersion of the models about the best-fit line is
$<0.03$~mag at redshifts $\lesssim 1.1$ and $\gtrsim 1.4$, and reaches
its largest value of 0.09~mag at $z=1.26$. We calculate the
$K$-correction for each galaxy using this best-fit relation,
effectively assuming that every galaxy is at the cluster
redshift. This results in an incorrect luminosity for non-cluster
member galaxies, but this is accounted for by performing the same
$K$-correction on the galaxies in the GOODS fields prior to
subtracting their luminosity.

\subsection{Luminosity function correction} \label{lumfaintgals}

We estimate the total luminosity of all galaxies below the detection
limit of $z_{850} = 24.72$ using a \citet{schechter76a}
luminosity function, which gives the number of galaxies in the
luminosity interval $[L,L+dL]$ in a given sample,
\begin{equation}
\Phi(L)dL = \Phi^\ast (L/L^\ast)^\alpha e^{-L/L^\ast} d(L/L^\ast).
\end{equation}
$\Phi^\ast$ is a normalization, $L^\ast$ is a
characteristic galaxy luminosity, and $\alpha$ is a unit-less
constant. The ratio of total to observed luminosity is then
\begin{equation}
C = \frac{\int_0^\infty L\Phi(L)dL}{\int_{L_{lim}}^\infty L\Phi(L)dL},
\end{equation}
and we multiply each observed cluster luminosity by C to get the total
luminosity. 

We assume values for $L^\ast$ and $\alpha$ determined in other studies
and use our data to perform a rough consistency check. For $\alpha$,
studies have shown that the value does not evolve much from low
redshift, at least for redder galaxies. Analyzing only red galaxies in
28 clusters spanning $0 < z < 1.3$, \citet{andreon08a} find $\alpha =
-0.91 \pm 0.06$ (rest-frame $V$-band) with no discernible trend in
redshift \citep[see also][]{andreon06a,andreon06c}. From five
intermediate-redshift clusters ($0.54<z<0.9$), \citet{crawford09a}
find a somewhat flatter faint-end slope $\alpha \sim -0.6$ (rest-frame
$B$-band) for the red-sequence luminosity function. Looking at the
full luminosity function, \citet{goto05a} find $\alpha = -0.82 \pm
0.10$ in one cluster at $z=0.83$ (rest-frame $B$-band), compared to
$\alpha = -1.00 \pm 0.06$ in 204 low-redshift clusters (rest-frame
$g$-band) \citep{goto02a}. In redder bands, \citet{strazzullo06a} find
$\alpha \sim -1$ for three clusters at redshifts $1.11 < z < 1.27$ (in
approximately rest-frame $z$ band). Summarizing, most studies find a
value consistent with $\alpha \sim -0.9$, and we assume this value in
computing $C$.

Values for $M^\ast$ are also reported in most of the above-mentioned
studies. Studies of red galaxies find that the variation of $M^\ast$
with redshift is consistent with passive evolution, with $M^\ast$
decreasing towards higher
redshifts \citep{andreon06c,crawford09a}. \citet{crawford09a} find
$M^\ast_B = -21.1$ and $M^\ast_B \sim -21.3$ (with errors of
approximately a half magnitude) for two clusters at redshifts 0.75 and
0.83. $K$-correcting from the observed
$[3.6]$-band, \citet{andreon06c} find $M^\ast_B \sim -21.7$ at $z \sim
1.1$, with approximately 0.5~magnitudes of evolution between $z=0.3$
and $z=1.1$. At lower redshift (considering all
galaxies) \citet{goto02a} find $M^\ast_B \sim -21.6$, compared to
$M^\ast_B \sim -21.0$ for one cluster at $z=0.83$ \citep{goto05a}. On
the basis of these measurements, we assume a value of $M^\ast_B =
-21.7$.

We have checked our assumed $M^\ast_B$ and $\alpha$ for consistency with
our data. With the set of spectroscopically-confirmed cluster galaxies
from our clusters at $z<1.2$, we confirmed that the bright end of the
luminosity function is consistent with $M^\ast_B = -21.7$, and strongly
inconsistent with values outside the range $M^\ast_B = -21.7 \pm
0.5$. We also determined the luminosity function using a statistical
subtraction of the ``background'' luminosity function from the GOODS
fields, finding excellent agreement with the assumed $M^\ast_B$ and
$\alpha$ values over the range $-24 < M_B < -19.8$ ($M_B = -19.8$
corresponds to the detection limit in the highest-redshift
clusters).

For each cluster, we calculate $C$ in the observer frame, converting
$M^\ast_B = -21.7$ to the observed $z_{850}$ band, using the cluster
redshift and a $K$-correction based on a passive galaxy template. In
Table~\ref{table:lum_params} we report the value $z_{850}^\ast$ and
the resulting correction $C$ for each cluster. The correction is less
than $5\%$ for the majority of clusters, rising to a maximum of 14\%
for the highest-redshift cluster. Because the correction is so small,
varying the assumed values of $M^\ast_B$ and $\alpha$ does not have a
large effect on the total luminosity. Varying $M^\ast_B$ by $\pm
0.5$~mag (a larger range than that allowed by our data) changes the
average correction by only $^{+4}_{-2}\%$. Varying $\alpha$ by $\pm
0.2$ changes the average correction by $^{+5}_{-2}\%$. We
conservatively take $^{+10}_{-3}\%$ (the full range when varying both
concurrently) as the systematic uncertainty in luminosity from the
faint-end correction (summarized in \S\ref{resultssys}).

\begin{deluxetable}{lccccc}
\tablewidth{0pt}
\tabletypesize{\scriptsize}
\tablecaption{\label{table:lum_params} Bright cutoff magnitudes 
and luminosity function parameters}
\tablehead{\colhead{ID} & \colhead{$z$} & \colhead{Cutoff from} 
& \colhead{$z_{850}^{\rm bright}$}
& \colhead{$z_{850}^\ast$} & \colhead{$C$}}
\startdata
A & 1.46 & Max cD & $21.09$ & $22.80$ & $1.143$\\
B & 1.12 & cD & $20.11$ & $21.38$ & $1.033$\\
C & 0.97 & cD & $19.87$ & $20.79$ & $1.018$\\
D & 1.02 & BCG & $20.13$ & $20.95$ & $1.021$\\
E & 1.03 & cD & $19.40$ & $20.99$ & $1.022$\\
F & 1.11 & Max cD & $19.63$ & $21.34$ & $1.031$\\
G & 1.26 & BCG & $20.34$ & $22.04$ & $1.064$\\
H & 1.24 & BCG & $20.33$ & $21.95$ & $1.058$\\
I & 1.34 & Max cD & $20.66$ & $22.37$ & $1.092$\\
J & 1.37 & Max cD & $20.77$ & $22.50$ & $1.104$\\
K & 1.41 & Max cD & $20.92$ & $22.65$ & $1.122$\\
L & 1.37 & Max cD & $20.77$ & $22.50$ & $1.104$\\
M & 0.90 & Max cD & $18.69$ & $20.53$ & $1.014$\\
N & 1.03 & BCG & $20.22$ & $20.99$ & $1.022$\\
P & 1.1\phn & Max cD & $19.58$ & $21.29$ & $1.030$\\
Q & 0.95 & cD & $20.01$ & $20.66$ & $1.015$\\
R & 1.22 & Max cD & $20.15$ & $21.86$ & $1.054$\\
S & 1.07 & Max cD & $19.44$ & $21.16$ & $1.026$\\
T & 0.97 & Max cD & $19.00$ & $20.75$ & $1.017$\\
U & 1.04 & Max cD & $19.31$ & $21.04$ & $1.022$\\
V & 0.90 & cD & $18.89$ & $20.49$ & $1.013$\\
W & 1.26 & Max cD & $20.33$ & $22.04$ & $1.064$\\
X & 1.10 & Max cD & $19.58$ & $21.34$ & $1.031$\\
Y & 1.24 & cD & $20.29$ & $21.90$ & $1.056$\\
Z & 1.39 & cD & $20.85$ & $22.58$ & $1.112$

\enddata
\tablecomments{``Cutoff from'' refers to how $z_{850}^{\rm bright}$ is determined.
``cD'': magnitude of visually central dominant galaxy. ``BCG'':
magnitude of visually classified brightest cluster elliptical (but not
central) galaxy. ``Max cD'': Cluster does not have obvious cD galaxy
or clear BCG. In this case, $z_{850}^{\rm bright}$ is $K$-corrected from
$M_B = -23.42$, the absolute magnitude of the brightest cD galaxy in
the entire sample.}
\end{deluxetable}

\subsection{Cluster Luminosities and Aggregate Cluster Profile}
\label{lumprofile}

For each cluster we sum the $K$-corrected $B$-band luminosity of all
galaxies brighter than the detection limit $z_{850} = 24.72$.  To
reduce noise, we discard galaxies that are clearly too bright to be
cluster members. In clusters with a central dominant (cD) galaxy or
dominant (but not central) brightest cluster galaxy(BCG), the bright
cutoff magnitude is set to the magnitude of the cD galaxy or BCG. In
clusters lacking a clearly dominant galaxy, we conservatively set the
cutoff based on the absolute magnitude of the most luminous cD galaxy
in any cluster, $M_B = -23.42$ (from cluster XMMU J2235.3$-$2557). The
bright cutoff magnitude in the observer frame, $z_{850}^{\rm bright}$,
is listed for each cluster in Table~\ref{table:lum_params}.  Because
the bright cutoff is chosen so conservatively, we expect that no
cluster galaxies are discarded. The effect of being overly
conservative is only to add noise, and this is captured in the
statistical uncertainty described below.

For each cluster we apply the same selection criteria and
$K$-corrections to the GOODS fields to determine the ``background''
specific to that cluster. The error in the luminosity comes from the
error in this background determination, which we estimate in the
following way: We select 30 non-connected circular regions (15 in each
of GOODS North and South) of radius $1.4'$, similar to the size of the
cluster fields. We determine the luminosity density in each of these
fields. The average is taken as the background luminosity for the
cluster, and the standard deviation (typically 15 -- 20 \% of the
average) is taken as the error in this ``background'' luminosity due
to variations between fields.

We have implicitly assumed that the GOODS average accurately
represents the cosmic average. GOODS incorporates only two widely
separated fields. As a result, the average luminosity density may
differ from the cosmic average due to variations in large scale
structure. As a rough estimate of the cosmic variance, we compare the
two GOODS fields.  The average luminosity density of the GOODS-North
regions is consistently higher than that of the GOODS-South regions by
15 -- 20\%. This means that the ``standard deviation'' of these two
samples of large scale structure is $\sim$8\%. We checked this using
the cosmic variance calculator made available
by \citet{trenti08a}\footnote{\url{http://casa.colorado.edu/~trenti/CosmicVariance.html}}. The
expected cosmic variance in galaxy number counts in the redshift
window $0.7 < z < 1.7$ for one GOODS field is approximately $\sim$6\%,
in good agreement with our na\"ive estimate. Conservatively, we take
$8\%$ as the cosmic variance for one GOODS field. For
the \emph{average} of the North and South fields, this implies a
cosmic variance of $8\%/\sqrt{2} \sim 6\%$.

One might be additionally concerned that the ``background'' in the
cluster fields is biased higher than the cosmic average because
clusters form in regions of large-scale overdensities. However, each
cluster field is a ``pencil-beam'' galaxy survey, so the vast majority
of non-cluster galaxies will not be associated with the high-density
region in which each cluster formed.

\begin{figure*}
\epsscale{1.175}
\plotone{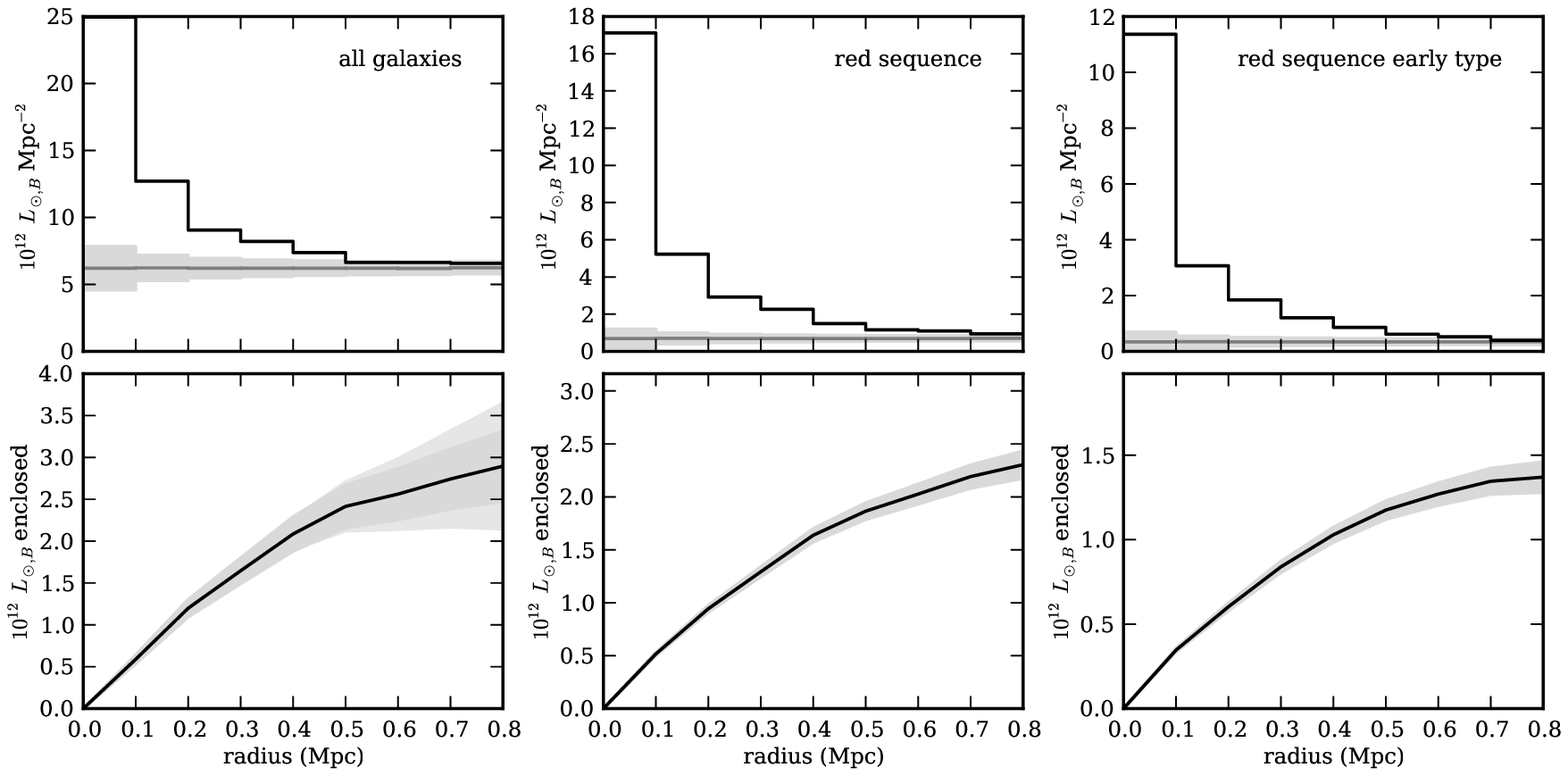}
\caption{Average luminosity profile of the 25 clusters. {\it Top row:}  
Average luminosity density in the cluster fields in annuli of width
0.1~Mpc extending out from the cluster center. The grey line and
shaded region show the estimated ``background'' luminosity in each
annulus and the error on that background, respectively. The darker
grey region is the statistical-only error, while the light grey is the
statistical $+$ cosmic variance error, added in quadrature. {\it
Bottom row:} The total enclosed luminosity as a function of radius,
derived by subtracting the background from the total luminosity
density in each bin in the top row plot. The left plots include
galaxies of all colors and morphologies, while the center plots
include only galaxies with $i_{775}-z_{850}$ colors within $\pm
0.2$~mag of the red sequence in their respective clusters. The right
plots include only galaxies that satisfy the color requirement and
also have $z_{850} < 24$ and are morphologically early type. By
excluding bluer galaxies (center and right plots) the background (and
error) is reduced dramatically.\label{fig:avgprofile_mult}}
\end{figure*}

Ideally one would measure a two-dimensional luminosity density,
$L(x,y)$, for each cluster, as in Equation~(\ref{eq:ratedenom}).
However, the large background makes this difficult. For our purpose
(which is to account for variations in control time with radius), it
is sufficient to assume the clusters have a circularly symmetric
luminosity distribution, $L(r)$. For each cluster, we sum the total
luminosity in annuli of width 0.1~Mpc. For nearly all clusters there
is a clear overdensity relative to the background out to
$r \sim 0.3$~Mpc. Beyond $0.3$~Mpc, the luminosity measurement is dominated
by background noise for most clusters. This might appear to be a
problem; we wish to characterize the cluster luminosities out to
$r \gtrsim 0.7$~Mpc, the area over which we searched for SNe. In fact,
it is only necessary to accurately measure the \emph{average}
luminosity profile over the full area (the denominator of
Eq.~\ref{eq:rate} is the sum of the cluster luminosities, weighted by
control time). Averaging all 25 clusters, there is a significant
measurement of the luminosity profile out to $>0.5$~Mpc
(Fig.~\ref{fig:avgprofile_mult}, left panels), and the average cluster
luminosity within $r<0.6$~Mpc has an error of $12\%$ (statistical
only) and $\sim 20\%$ (statistical $+$ cosmic variance), below the
Poisson error in the number of SNe detected.

Beyond $r<0.6$~Mpc, the control time is generally small (that is,
there are few observations covering the outskirts of the clusters) and
the cluster luminosity density is low, meaning that these regions will
not contribute greatly to the rate measurement. Still, we include
these regions in our rate calculation, using the entirely reasonable
prior that the luminosity density is decreasing with radius past
$r<0.6$~Mpc. How rapidly the luminosity density decreases will not
have a significant impact on the result, but as a convenient analytic
description we fit a $\beta$-model of the form
\begin{equation}
L(r) = \frac{\Sigma_0}{(1+(r/r_{\rm core})^2)^\beta}
\end{equation}
over the range $r<0.6$~Mpc and apply this function at $r>0.6$~Mpc. The
data are well-fit by this model, with best-fit parameters $r_{\rm
core} = 0.074$~Mpc and $\beta= 0.91$. Varying this model luminosity by
$\Delta\Sigma_0 = \pm 20\%$ (easily enclosing the allowed range of
$L(r)$) only changes our results by $\pm 4\%$. This and other
systematic uncertainties are summarized in
Table~\ref{table:clratesys}.

\begin{deluxetable*}{lccccc}
\tablewidth{0pt}
\tabletypesize{\scriptsize}
\tablecaption{\label{table:lum_avg} Average cluster luminosities within $r < 0.6$~Mpc}
\tablehead{\colhead{Cluster subset} & $N_{\rm clusters}$ & \colhead{$\bar{z}$} 
& \colhead{All galaxies ($10^{12} L_{\odot,B}$)}
& \colhead{RS galaxies ($10^{12} L_{\odot,B}$)}
& \colhead{RSE galaxies ($10^{12} L_{\odot,B}$)}}
\startdata
X-ray discovered & 9 & 1.20 & $ 2.86 \pm  0.54 \pm  0.45$ & $ 2.42 \pm  0.16 \pm  0.05$ & $ 1.47 \pm  0.12 \pm  0.02$ \\
IR-Spitzer discovered & 7 & 1.30 & $ 2.85 \pm  0.70 \pm  0.52$ & $ 1.83 \pm  0.24 \pm  0.07$ & $ 0.96 \pm  0.16 \pm  0.03$ \\
Optical discovered & 9 & 1.00 & $ 1.99 \pm  0.37 \pm  0.32$ & $ 1.75 \pm  0.08 \pm  0.03$ & $ 1.29 \pm  0.06 \pm  0.01$ \\[0.1in]
$z<1.2$ & 14 & 1.02 & $ 2.14 \pm  0.31 \pm  0.33$ & $ 1.79 \pm  0.07 \pm  0.03$ & $ 1.28 \pm  0.05 \pm  0.01$ \\
$z>1.2$ & 11 & 1.32 & $ 3.06 \pm  0.58 \pm  0.54$ & $ 2.31 \pm  0.19 \pm  0.07$ & $ 1.23 \pm  0.14 \pm  0.04$ \\[0.1in]
All Clusters & 25 & 1.15 & $ 2.54 \pm  0.31 \pm  0.42$ & $ 2.02 \pm  0.09 \pm  0.05$ & $ 1.26 \pm  0.07 \pm  0.02$ 

\enddata
\tablecomments{``RS'': galaxies within $\pm 0.2$~mag
of the cluster red sequence. ``RSE'': galaxies fulfilling the ``RS''
requirement, and also $z_{850} < 24$, and morphologically
early-type. The first and second confidence intervals are the
statistical error and cosmic variance error, respectively. These
luminosities do not include the faint-galaxy correction $C$.}
\end{deluxetable*}

\subsection{Galaxy subsets} \label{lumsubsets}

In addition to measuring the total luminosity of all galaxies in the
clusters, we also measure the total luminosity of only red-sequence
galaxies and the total luminosity of only red-sequence, morphologically
early-type galaxies. These measurements enable us to compute the
cluster SN~Ia rate specifically in these galaxy subsets. For the
red-sequence-only measurement we follow the same procedure as above,
but eliminate from the analysis all galaxies with $i_{775} - z_{850}$
colors more than 0.2~mag from their respective cluster red sequences
(galaxy colors and cluster red sequences are determined as in
Meyers11). For the red-sequence early-type measurement, we make the same
requirement in color, and additionally use the quantitative morphology
requirements of Meyers11. Meyers11 use two parameters, asymmetry and Gini
coefficient, to automatically divide galaxies into early- and
late-type subsets. Here we require the asymmetry to be $<0.10$ and the
Gini coefficient to be $>0.40$. We also require the galaxies to be
$z_{850} < 24$ as the asymmetry and Gini coefficient are somewhat less
reliable at fainter magnitudes.

The luminosity profiles for these two subsets are shown in the center
and right columns of Figure~\ref{fig:avgprofile_mult}. The profiles
are broadly consistent with the profile of the full cluster luminosity
(left column), but the ``subset'' profiles are much better
measured. This is because by excluding bluer galaxies, we have
eliminated much of the background while still retaining the majority
of cluster galaxies. The red-sequence subset contains $77\%$ of the
luminosity of the full cluster within $0.6$~Mpc
(Table~\ref{table:lum_avg}).  The red-sequence early-type subset has
$62\%$ of the light contained in the red-sequence subset. However,
keep in mind that in the early-type subset we have excluded
$z_{850}>24$ galaxies, whereas they are included in the red-sequence
subset: In fact $68\%$ of $z_{850}>24$ red-sequence galaxies pass the
``early-type'' morphology requirements.

Note that our definition of ``red-sequence'' here is a relatively
simple one. It is sufficient to select a subsample of ``more red''
galaxies for the purpose of looking for a dependence of the SN rate
with galaxy color within the cluster. However, for measuring the red
fraction in clusters
\citep[e.g., the Butcher-Oemler effect][]{butcher78a,butcher84a}, 
defining red-sequences with a constant color width for all redshifts
is not ideal \citep{andreon06d}. The luminosity content of the subsets
are reported above only to give the relative size of each sample; a
full analysis of the cluster content is beyond the scope of this
paper.

\subsection{Stellar Mass-to-Light Ratio} \label{lummass}

To compare SN rates in clusters of different ages, rate measurements
must be normalized by stellar mass rather than stellar luminosity because
luminosity changes as stars age. To convert our luminosity
measurements to mass measurements we use a mass-to-light ($M/L$) ratio
based on a stellar evolution model. There are several available models
in the literature. The choice of stellar tracks, metallicity, star
formation history, and in particular the assumed IMF, will all affect
the derived $M/L$ ratio to some extent. For the purpose of measuring
the change in rate with redshift, it is important to use
a \emph{consistent} model and assumptions for determining the $M/L$ ratio for all rate measurements. That
is, we are most concerned that the model accurately captures the
evolution of stellar luminosity over the redshift range of interest
($0<z<1.46$), and less concerned about the overall normalization of the $M/L$ ratio. To that end, for our main result we will use a model and
assumptions that match as closely as possible those used for the $M/L$ ratio in 
low-redshift cluster rate measurements. As we also give results
normalized by luminosity, those wishing to use a different $M/L$ ratio
can easily do so. Finally, note that the \emph{initial} stellar mass formed is
the quantity of interest for normalizing rate measurements. However,
as most rate measurements and $M/L$ ratios have been reported in terms
of current mass, we give our results in these units and simply note
the difference between current and initial mass for the purpose of
comparing rate measurements. Thus, in the following paragraphs $M$ refers to
current stellar mass.

\subsubsection{$M/L$ ratio in low-redshift cluster rate measurements}

The lower-redshift cluster rate studies
of \citet{sharon07a}, \citet{sharon10a}, and by
extension, \citet{dilday10a} have used the relations between $M/L$
ratio and galaxy color derived by \citet[][hereafter Bell03]{bell03a}.
For example, \citet{sharon07a} use the relation $\log_{10} (M/L_z) =
-0.052 +0.923(r-i)$ and \citet{sharon10a} use $\log_{10} (M/L_g) =
-0.499 +1.519(g-r)$, where $M$, $L_z$ and $L_g$ are in solar units.
In order to use a consistent model, it is important to recognize how
these relations were derived.  Bell03 fit a grid of {\sc
p{\'e}gase2} \citep{fioc97a} synthetic galaxy spectral energy
distributions (SEDs) to actual $ugrizK$ photometry of low-redshift
galaxies. The grid covers a range of metallicities and star formation
histories. The star formation histories have exponentially-decreasing
or -increasing star formation rates, and assume that star formation
commenced at $z=4$. For each galaxy, the $M/L$ ratio is that of the
best-fit synthetic galaxy SED, consistently evolved to $z=0$.  Bell03
use a ``diet'' \citet{salpeter55a}
IMF \citep[following][]{bell01a}. This IMF is defined as having the
same colors and luminosity as a Salpeter IMF, but with a total mass
30\% lower. The difference in mass is attributed to a smaller number
of faint low-mass stars relative to a Salpeter IMF. These stars do
not contribute significantly to the luminosity of the Salpeter
IMF. The diet Salpeter IMF results in $M/L$ ratios 30\% lower at a
given color than a normal Salpeter IMF. Note that because Bell03
simply take the $M/L$ ratio from the best-fit synthetic SED of each
galaxy, the Bell03 relations will generally fall within the grid of
$M/L$ versus color covered by the synthetic galaxy SEDs.

\begin{figure*}
\epsscale{1.175}
\plotone{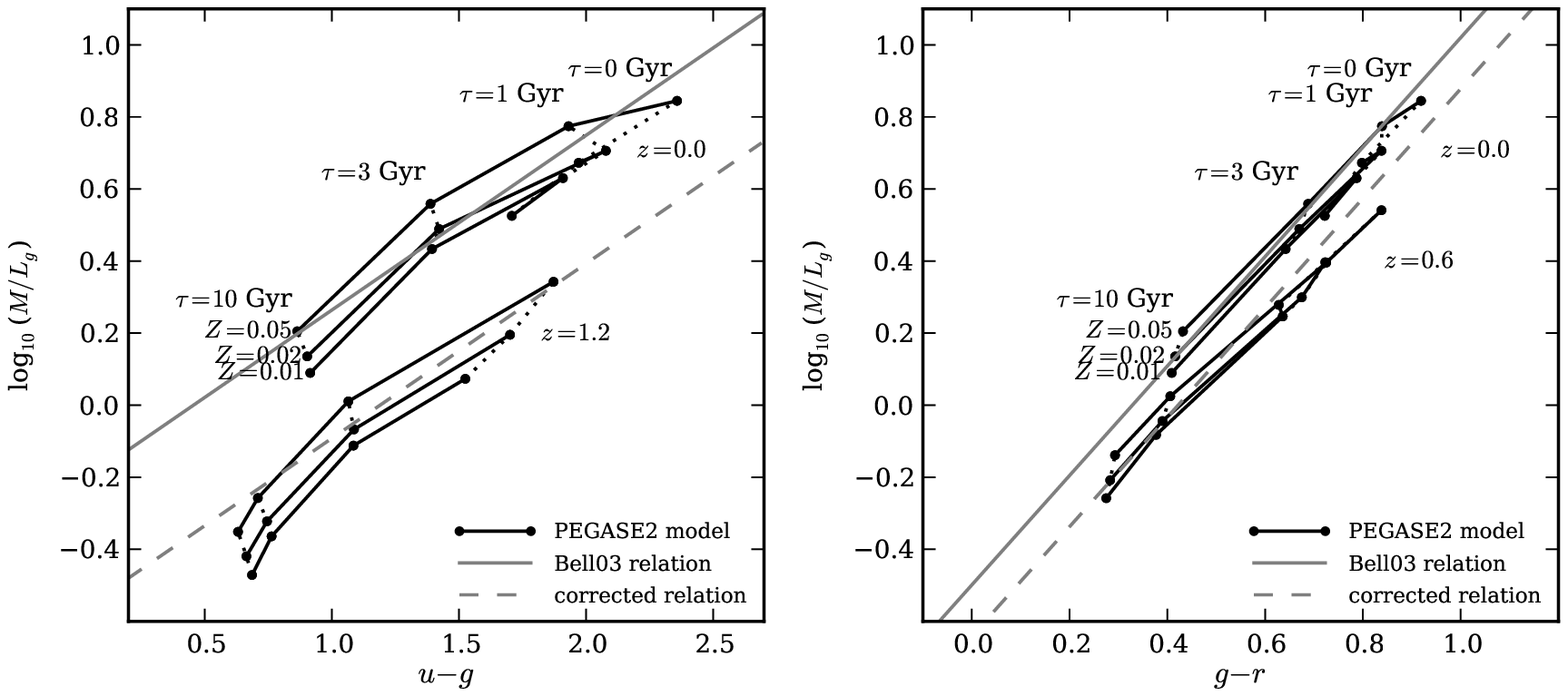}
\caption{Evolution of $M/L$ ratio versus color with redshift. 
\emph{Left panel:} $M/L$ ratio as a function of $u-g$ color 
at $z=0$ and at $z=1.2$ (typical redshift in this study). The grid of
points show {\sc p{\'e}gase2} models with exponentially-decreasing
star formation rates with e-folding times $\tau$ and metallicities
$Z$. For each model, star formation begins at $z=4$. Models with
constant metallicity are connected by solid black lines and models
with identical star formation histories are connected by dotted
lines. For example, models with $\tau = 0$, corresponding to a simple
stellar population, are the rightmost points (corresponding to
$Z=0.01$, $0.02$, $0.05$) connected by dotted lines. As the models are
evolved back in time from an observed redshift of $z=0$ to an observed
redshift of $z=1.2$, the $M/L$ ratio decreases and moves away
from the Bell03 relation (\emph{solid grey line}). The \emph{dashed
grey line} shows the relation used in this study for $z=1.2$. At
$z=1.2$ the offset from the Bell03 relation is $-0.36$~dex, or a
factor of 0.43. \emph{Right panel:} Same as left panel, but for $g-r$
color and for an observed redshift of $z=0.6$, the typical redshift in
the rate study of \citet{sharon10a}. The offset here is only
$-0.14$~dex, or a factor of 0.72.
\label{fig:mlratio}}
\end{figure*}

\subsubsection{$M/L$ ratio at $0.9<z<1.46$}

Ideally, for consistency with \citet{sharon07a}, \citet{sharon10a}
and \citet{dilday10a}, we would simply use the Bell03 relation for
$u-g$ color, which most closely matches our observed color: $\log_{10}
(M/L_g) = -0.221 +0.485(u-g)$. However, the Bell03
relations are based on $ugrizK$ photometry of low-redshift galaxies,
corrected for evolution to $z=0$. As such, they are specific to $z=0$
and not directly applicable at high redshift. A stellar population
passively evolving from age a few~Gyr (at $z \sim 1$) to $> 10$~Gyr
(at $z=0$) will dim significantly while only growing slightly redder
(see, e.g. BC03), in a manner that does not follow the Bell03
relations.  To estimate the effect of evolution from their $z=0$
relation to higher redshift, we make a similar grid of {\sc
p{\'e}gase2}-generated SEDs with the same formation redshift,
metallicities, IMF, and star formation histories. As expected, when
evaluated at $z=0$, the $M/L$ ratios of this grid are consistent with
the Bell03 relation (Fig.~\ref{fig:mlratio}, left panel, upper grid of
black points). Evaluating the SEDs at higher redshifts, we find that
the $M/L$ ratios are well fit by a relation with the same slope, but
smaller normalization. For example, at $z=1.2$, the best-fit offset
from the $z=0$ relation is $-0.36$~dex (Fig.~\ref{fig:mlratio}, left
panel, dashed line). At the extremes of the redshift range of
interest, the best fit offset is $-0.26$~dex ($z=0.9$) and $-0.44$~dex
($z=1.46$). We therefore use a $M/L$ ratio of
\begin{equation} \label{eq:mlratio}
\log_{10} (M/L_g) = \left\{ \begin{array}{ll}
-0.48+0.485(u-g), & z=0.9 \\
-0.66+0.485(u-g), & z=1.46
\end{array} \right.
\end{equation}
and linearly interpolate for intermediate redshifts. Another way to
view Equation~(\ref{eq:mlratio}) is that, independently of the
relation at $z=0$, we have fit a linear relation to the {\sc
p{\'e}gase2} SEDs at the redshift of each cluster, assuming a slope
consistent with Bell03.

Using Equation (\ref{eq:mlratio}) we calculate mass on a
galaxy-by-galaxy basis: we $K$-correct the observed $i_{775}$ and
$z_{850}$ magnitude to rest-frame SDSS $u$ and $g$ magnitudes using
the method discussed in \S\ref{lumkcorrections}, and obtain the $M/L$
ratio from the $u-g$ color. In all, 66\% of the clusters'
luminosity is from galaxies with color in the range $1.3 < u-g < 1.7$,
27\% of the luminosity is distributed roughly equally between galaxies
in the range $0.6 < u-g < 1.3$, and the remainder is in redder
galaxies with $u-g > 1.7$. Thus, while there is a clear presence of
bluer cluster galaxies, the majority of the clusters luminosity is
confined to a narrow range in color. This narrow color range means
that changes in the assumed slope of Equation (\ref{eq:mlratio}) will
not have a large effect on the resulting total mass.

The cumulative $M/L$ ratio (the ratio of the total mass of all 25
clusters to the total luminosity of all 25 clusters) is $M/L_g = 1.25$
(see Table~\ref{table:clrates}, ``denom''). For red-sequence galaxies
only, the ratio is higher ($M/L_g = 1.38$) due to the exclusion of
bluer galaxies with a lower inferred $M/L$ ratio.

\subsubsection{$M/L$ ratio uncertainty} \label{lummasserror}

As noted above, we are primarily concerned with the accuracy of the
evolution in the stellar mass and luminosity over the range
$0<z<1.46$, rather than the accuracy of the absolute $M/L$ ratio.  As
a cross-check of the $M/L$ ratio evolution, we have compared the above
results (using {\sc p{\'e}gase2}) to the results obtained with the
BC03 SEDs. We use the standard Padova 1994 evolution and the same star
formation histories as above. In terms of evolution offset from $z=0$
to $z \sim 1.2$, we find results consistent within 0.03~dex.

This consistent evolution in BC03 and {\sc p{\'e}gase2} is
encouraging. However, to be much more conservative in our estimate of
the uncertainty in the $M/L$ ratio evolution, we take the scatter of
the models around the best-fit line as our uncertainty. In
Figure~\ref{fig:mlratio}, in the color range of interest, the scatter
is approximately $\pm 0.08$~dex (20\%) at both low and high
redshift. We use this as the systematic uncertainty in the $M/L$ ratio
for the purpose of comparing SN rates at low and high redshift
in \S\ref{conclusionsdtd} and \S\ref{conclusionssys}. The uncertainty
in the absolute $M/L$ ratio is much greater, due mainly to the
uncertainty in the true IMF.

\section{Results and Systematic Uncertainties} \label{results}

Here we present our results for the full cluster rate and for two
galaxy subsets (\S\ref{resultsresults}) and summarize contributions to
the uncertainty (\S\ref{resultssys}) in each. In \S\ref{ratevscut} we
show that the rate result in the subsets are not sensitive to the
specific parameters used to select the subset.

\subsection{Results} \label{resultsresults}

\begin{deluxetable*}{lcccccccccc}
\tablecaption{\label{table:clrates}Results}
\tablehead{\colhead{Environment} & \colhead{Unit} & \colhead{$\bar{z}$} &
   \colhead{$N_{\rm SN~Ia}$} & \colhead{Denom} & \colhead{Rate} & 
   \colhead{(stat)} & \colhead{(sys)}}
\startdata
Full cluster & SNuB & 1.14 & $8.0 \pm 1.0$ & 15.87 & 0.50 & $^{+0.23}_{-0.19}$ & $^{+0.10}_{-0.09}$\\
Full cluster & SNug & \nodata & \nodata & 15.96 & 0.50 & $^{+0.23}_{-0.19}$ & $^{+0.10}_{-0.09}$\\
Full cluster & SNuM & \nodata & \nodata & 22.41 & 0.36 & $^{+0.16}_{-0.13}$ & $^{+0.07}_{-0.07}$\\
Red-sequence & SNuB & 1.13 & $6.5 \pm 0.5$ & 11.95 & 0.54 & $^{+0.25}_{-0.19}$ & $^{+0.07}_{-0.07}$\\
Red-sequence & SNug & \nodata & \nodata & 12.20 & 0.53 & $^{+0.24}_{-0.19}$ & $^{+0.07}_{-0.07}$\\
Red-sequence & SNuM & \nodata & \nodata & 17.61 & 0.37 & $^{+0.17}_{-0.13}$ & $^{+0.05}_{-0.05}$\\
Red-sequence early-type & SNuB & 1.10 & $6.0 \pm 0.0$ &  7.29 & 0.82 & $^{+0.39}_{-0.30}$ & $^{+0.09}_{-0.08}$\\
Red-sequence early-type & SNug & \nodata & \nodata &  7.59 & 0.79 & $^{+0.38}_{-0.29}$ & $^{+0.09}_{-0.08}$\\
Red-sequence early-type & SNuM & \nodata & \nodata & 11.77 & 0.51 & $^{+0.24}_{-0.19}$ & $^{+0.06}_{-0.05}$

\enddata
\tablecomments{``Denom'' is the denominator of equation~(\ref{eq:rate}) 
and has units of $10^{12} L_{\odot,B}$~years, $10^{12} L_{\odot,g}$~years and 
$10^{12} M_\odot$~years for rate units of SNuB, SNug and SNuM
respectively.}
\end{deluxetable*}

The results are presented in Table~\ref{table:clrates}.  We derive a
rate in the full cluster, in red-sequence galaxies only, and in
red-sequence early-type galaxies only. Each subset includes a
different number of SNe: As discussed in \S\ref{candsummary}, we have
discovered $8 \pm 1$ cluster SNe, where the quoted uncertainty is due
to classification uncertainty (including uncertainty in both SN type
and cluster membership).  Limiting the sample to only SNe discovered
in galaxies included in the red-sequence subset excludes SN~SCP06F12
and SN~SCP06C1, leaving $6.5 \pm 0.5$ cluster
SNe~Ia. The uncertainty here comes from the uncertainty in the cluster
membership and type of SN~SCP06E12, which we count $0.5 \pm 0.5$
cluster SNe~Ia.  Further limiting the sample to only SNe discovered in
galaxies included in the red-sequence early-type subset, SN~SCP06E12
is eliminated as its host galaxy is dimmer than the $z_{850} = 24$
cutoff used for this subset leaving $6$ SNe~Ia with negligible
classification error. The number of SNe~Ia discovered in each subset,
including classification error, is summarized in
Table~\ref{table:clrates} under $N_{\rm SN~Ia}$.

We normalize the rate in three different ways: by $B$-band luminosity,
by $g$-band luminosity, and by stellar mass.  For each cluster, we use
the visibility time map $T(x,y)$ (e.g., Fig.~\ref{fig:ctmaps}) and the
measured luminosity (or mass) profile to carry out the integral in
equation~(\ref{eq:ratedenom}) giving the time-luminosity searched. The
sum of these values for all 25 clusters is the denominator of
equation~(\ref{eq:rate}), the total time-luminosity searched in all
clusters. This is shown in Table~\ref{table:clrates} under ``Denom''
for each sample. The rate is simply $N_{\rm SN~Ia}$ divided by
``denom,'' as in equation~(\ref{eq:rate}). The contributions to the
statistical and systematic errors are summarized in
Table~\ref{table:clratesys}.

The weighted-average redshift, $\bar{z}$, for each subsample is given by
\begin{equation}
\bar{z} = \frac{\sum_i z_i \int_{x,y} T_i(x,y) L_i (x,y)}
        {\sum_i \int_{x,y} T_i (x,y) L_i (x,y)},
\end{equation}
where $z_i$, $L_i$ and $T_i$ are the redshift, luminosity and
effective visibility time of the $i$-th cluster, respectively. The
weighted-average redshift is slightly smaller for the red-sequence and
red-sequence early-type galaxy subsets. This is because in the
higher-redshift clusters, a smaller fraction of galaxies meet the
subset requirements (see $z<1.2$ versus $z>1.2$ average cluster
luminosity in Table~\ref{table:lum_avg}).

\subsection{Summary of Systematic Uncertainties} \label{resultssys}

\begin{table} \label{table:clratesys}
\tablewidth{\columnwidth}
\caption{\label{table:clratesys}Sources of Uncertainty}

\begin{tabular}{l c c c}
\hline
\hline
                &  Full    &   Red-    & Red-sequence \\ 
                & cluster  & sequence  & early-type   \\
Source of error &  (\%)    &   (\%)    &     (\%)     \\
\hline
\hline
\multicolumn{4}{c}{Statistical} \\
\hline
Poisson                 & $^{+40}_{-32}$ & $^{+45}_{-35}$ & $^{+47}_{-36}$\\
Luminosity (stat)       & $\pm 12$      & $\pm 6$       & $\pm 6$     \\
Luminosity (cosmic var.)& $\pm 16$      & $\pm 4$       & $\pm 3$     \\[0.1in]
\bf{Total statistical}  & $^{+45}_{-38}$ & $^{+46}_{-35}$ & $^{+48}_{-37}$\\
\hline
\multicolumn{4}{c}{Systematic} \\
\hline
SN type classification  & $\pm 13$      & $\pm 8$       & \nodata    \\
Control time: varying $M_B$& $^{+8}_{-6}$   & $^{+8}_{-6}$& $^{+8}_{-6}$ \\
Control time: dust distribution & $^{+10}_{-2}$ & \nodata & \nodata     \\
Luminosity: MAG\_AUTO corr.    & $\pm 7$       & $\pm 7$   & $\pm 7$\\
Luminosity: $K$-correction          & $\pm 3$       & $\pm 3$   & $\pm 3$\\
Luminosity: Faint galaxy corr. & $^{+4}_{-9}$   & \nodata   & \nodata\\
Luminosity: $r>0.6$(0.8)~Mpc  & $\pm 4$   & $\pm 1$    & $\pm 1$\\[0.1in]
\bf{Total systematic} & $^{+20}_{-19}$ & $^{+14}_{-12}$  & $^{+11}_{-10}$\\[0.05in]
\hline 
\\
\bf{Total statistical $+$ systematic} & $^{+49}_{-42}$ & $^{+48}_{-37}$ & $^{+49}_{-38}$ \\
\hline
\end{tabular}
\end{table}

Throughout the paper, we have highlighted and addressed possible
sources of systematic uncertainty. Here we summarize these sources.
In Table~\ref{table:clratesys} we show the relative contribution of
each to the total systematic error, and compare to sources of
statistical error.

(1) \emph{SN type classification:} The uncertainty in the number of
SNe observed in each galaxy subset was addressed in
\S\ref{resultsresults}. The fractional error in the rate is simply the
fractional error in the number observed.

(2) \emph{Control time: Varying $M_B$:} In our control time
simulations, we assumed a distribution of SN~Ia light curve shapes and
absolute magnitudes. To first order, the impact of these assumptions
on the control time is captured by varying the assumed SN~Ia absolute
magnitude (\S\ref{ctsys}). Variations of $\pm 0.2$~mag resulted in a
rate change of $^{+8}_{-6}\%$

(3) \emph{Control time: dust distribution:} In \S\ref{ctsys} we
assessed the impact of varying amounts of dust extinction on the
control time. Assuming an unrealistically large amount of
dust-affected SNe decreased the control time by 9\% (increasing the SN
rate by $10\%$), while decreasing the amount of dust-affected SNe
increased the control time by $2\%$ (decreasing the SN rate by
$2\%$). We do not apply this systematic error to the red-sequence or
red-sequence early-type subsets, as we have independent evidence that
the amount of dust is limited in these environments.

(4) \emph{MAG\_AUTO correction:} In computing the total $z_{850}$ luminosity
of each galaxy, we made a correction to the MAG\_AUTO magnitude
ranging from $\sim$10\% at $z_{850}=20$ to $\sim$30\% at
$z_{850}=25$. Varying the range of $n$ used in the simulation by $\pm
1$ affects the correction by $\pm 7\%$.

(5) \emph{$K$-correction:} In \S\ref{lumkcorrections}, we noted that the
scatter of BC03 templates about the best-fit $K$-correction is
typically less than 0.03~mag. We use this value as the systematic
error on the $K$-correction.

(6) \emph{Faint galaxy correction:} The average correction $C$
reported in Table~\ref{table:lum_params} is 1.054. Varying $M^\ast_B$ by
$\pm$ 0.5~magnitudes results in an average correction of 1.032 and
1.092 for $-0.5$ and $+0.5$~magnitudes, respectively. Varying $\alpha$
by $\pm 0.2$ results in an average correction of 1.027 and 1.098 for
$\alpha = -0.7$ and $-1.1$, respectively. Concurrently varying
$M^\ast_B$ and $\alpha$ within the same ranges results in a minimum
average correction of 1.015 ($M^\ast_B = -22.2$, $\alpha = -0.7$) and a
maximum average correction of 1.154 ($M^\ast_B = -21.2$, $\alpha =
-1.1$). Conservatively, we assign $^{+4\%}_{-9\%}$ as the systematic
error on the rate associated with this correction. This error is not
applied to the red-sequence or red-sequence early-type subsets because
these subsets do not include light from galaxies below the detection
threshold.

(7) \emph{Luminosity at large radii:} In \S\ref{lumprofile} we assumed
a model for the cluster luminosity profile at $r>0.6$~Mpc (0.8~Mpc for
red-sequence and red-sequence early-type subsets). Varying the model
luminosity by $\pm 20\%$ resulted in a $\pm 4\%$ change in the full
cluster rate. The change is much smaller ($\pm 1\%$) for the red
galaxy subsets because the model is only used at $r>0.8$~Mpc.

(8) \emph{$M/L$ ratio:} In \S\ref{lummass} we used a $M/L$ ratio to
translate stellar luminosity to stellar mass. Rather than estimating
the absolute uncertainty in the $M/L$ ratio (which is strongly
dependent on assumptions), we estimated the uncertainty in
the \emph{evolution} of the $M/L$ ratio from low to high
redshift. This is the relevant uncertainty for comparing rates at
differerent redshifts in order to derive the SN~Ia delay time
distribution. We defer discussion of this uncertainty
to \S\ref{conclusionssys} where we discuss uncertainties in the DTD.

\subsection{Effect of Varying Subset Requirements} \label{ratevscut}

\begin{figure}
\epsscale{1.175}
\plotone{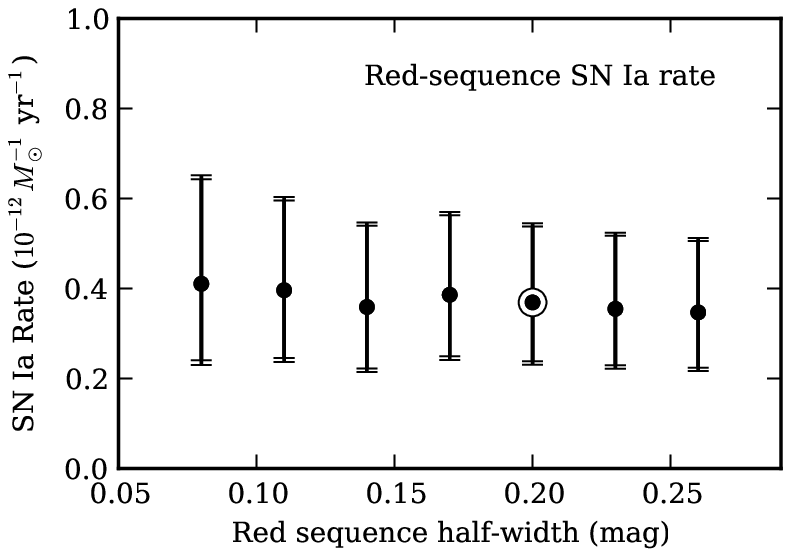}
\caption{The effect of varying the width of the red sequence. The
  nominal red-sequence rate result corresponds to a half-width of
  0.20~mag. The inner and outer error bars represent the statistical
  and total uncertainty, respectively.
\label{fig:ratevscut1}}
\end{figure}

\begin{figure}
\epsscale{1.175}
\plotone{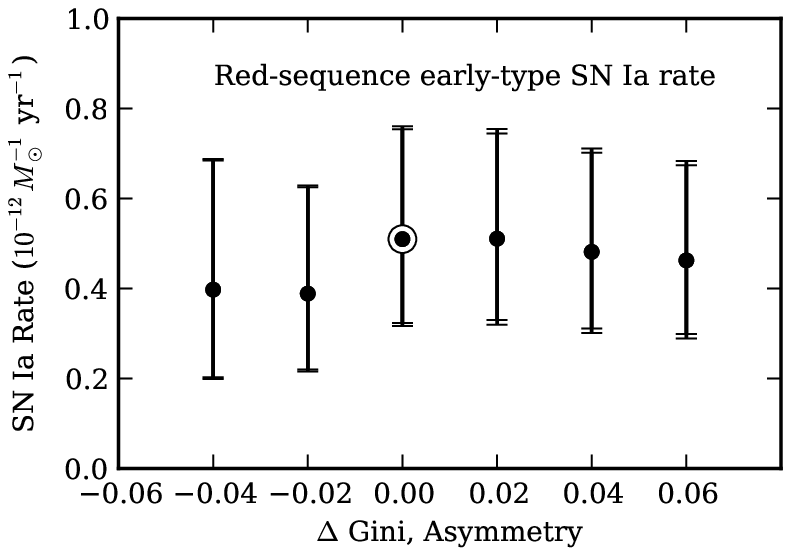}
\caption{The effect of varying the morphology parameter
  requirements. Negative $\Delta$ values correspond to a more strict
  selection and a higher-purity early-type galaxy sample. The
  requirements are asymmetry $<0.1+\Delta$ and Gini coefficient $>
  0.40-\Delta$. The nominal red-sequence early-type rate corresponds
  to $\Delta = 0$.  The red-sequence half-width is fixed at
  0.2~mag. The inner and outer error bars represent the statistical
  and total uncertainty, respectively.
\label{fig:ratevscut2}}
\end{figure}

In selecting our red-sequence and red-sequence early-type galaxy
subsamples, we required red-sequence galaxies to be within $\pm
0.2$~mag of the color of their cluster red sequence. For early-type
galaxies, we required the asymmetry parameter to be $< 0.1$ and the
Gini coefficient to be $> 0.40$. It is interesting to test the
sensitivity of the results to variations in the requirements. In
Figures~\ref{fig:ratevscut1} and~\ref{fig:ratevscut2} we vary the
requirements and observe the effect on the rates. As requirements are
made more strict (for example, narrowing the red sequence) the total
mass of the sample decreases. At the same time, SNe fall out of the
sample when their host galaxies are cut. The Poisson error increases
as the number of included SNe shrinks.

There is not a strong dependence of the SN~Ia rate with galaxy color
residual from the red sequence (Fig.~\ref{fig:ratevscut1}). Even in
cluster galaxies that lie in a tight range around the red-sequence
($\pm 0.08$~mag), we find a SN~Ia rate consistent with the full
cluster rate. Similarly, there is no significant rate trend with the
purity of the early-type sample (Fig.~\ref{fig:ratevscut2}). We
happened to pick morphology requirements that yield a slightly higher
rate than other choices, but such variations are expected with
small-number statistics and are accounted for by the Poisson
uncertainty in the result (Tables~\ref{table:clrates}
and~\ref{table:clratesys}). Even in the most-selective subset ($\Delta
= -0.04$), the rate is consistent with the full cluster rate.

\section{Discussion and Conclusions} \label{conclusions}
\subsection{Host-less Cluster SNe~Ia}

As reported by Dawson09, we have discovered one potential host-less
cluster SN~Ia among the $8 \pm 1$ cluster SNe~Ia. SN~SCP06C1 is
projected near two possible host galaxies: A $z_{850} = 21.6$ spiral
galaxy $1''.1$ West of the SN, and a significantly fainter $z_{850} =
24.6$ galaxy $0''.45$ ($\sim$3.5~kpc at the cluster redshift)
Northeast of the SN (See Dawson09, Fig.~2).

The galaxy-subtracted SN spectrum clearly shows a SN~Ia at redshift
$z=0.98$ near maximum light, consistent with the light curve fit. The
redshift of $z=0.98 \pm 0.01$ is consistent with the cluster redshift
of 0.974.  The bright spiral galaxy is actually in the background of
the cluster, at $z=1.091$. Strong [O{\sc ii}] emission is visible in
the spectrum, along with Ca H \& K and H$\delta$
absorption. Unfortunately, the small separation between the main
galaxy and the smaller galaxy to the Northeast means that the spectrum
of the smaller galaxy is dominated by light from the larger galaxy,
making it impossible to assess a redshift. It is thus possible that
the small galaxy is at the cluster redshift and is the actual host of
the SN. Alternatively, the small galaxy might be at the same redshift
as the larger galaxy and physically associated with it (either as a
satellite galaxy or as part of the spiral structure of the galaxy). It
is interesting to note that the SN is only $20''$ (160~kpc) projected
radius from the center of the cluster, perhaps giving more weight to
the hypothesis that it is associated with a diffuse intracluster
stellar component.

Not being able to confirm or reject this SN as host-less, we have an
upper limit of one host-less SN out of a total of $8 \pm
1$. Discovering one host-less SNe~Ia out of seven total would imply an
intrinsic host-less SN~Ia fraction of $14\% ^{+18\%}_{-7\%}$ (binomial
$68\%$ confidence intervals), and a 95\% upper limit of $<47\%$. This
is broadly consistent with host-less SN~Ia constraints at intermediate
redshifts \citep{sharon10a} and at low redshift
\citep{galyam03a,sand10a}. At low redshift it has been possible to
confirm the host-less nature of some SNe using deeper follow-up
imaging, leading to better constraints. The upper limit of $<47\%$ is
also consistent with direct measurements of intracluster light at low
redshift, but does not strongly constrain evolution. A sample twice
the size or larger, with deeper follow-up to confirm host-less SNe~Ia
would begin to place interesting constraints on hypotheses for the
formation of the intracluster stellar component from $z>1$ to today.

\subsection{Comparison to Other Cluster Rate Measurements}

Cluster SN~Ia rates have been reported at lower redshifts by several
groups. In nearby ($z \lesssim 0.2$) clusters, measurements include
those of \citet{sharon07a} at $z \sim 0.14$, \citet{mannucci08a} at $z
\sim 0.02$, and \citet{dilday10a} at $z \sim 0.09$ and $z \sim 0.22$.
At intermediate redshifts, \citet{sharon10a} recently reported
the rate in $0.5 < z < 0.9$ clusters (median $z \sim 0.6$). At higher
redshifts, \citet{galyam02a} placed the first constraints on the $z
\gtrsim 0.8$ cluster rate using a sample of three clusters at
$z=0.83$, $0.89$ and $z=1.27$.  However, their SN sample included only
one firm SN~Ia at $z=0.83$. The resulting rate has correspondingly
large uncertainties and essentially places only an upper limit on the
$z>0.9$ cluster rate. Our result is thus a large step forward in the
measurement of the SN rate in the highest-redshift clusters.

\begin{figure}
\epsscale{1.175}
\plotone{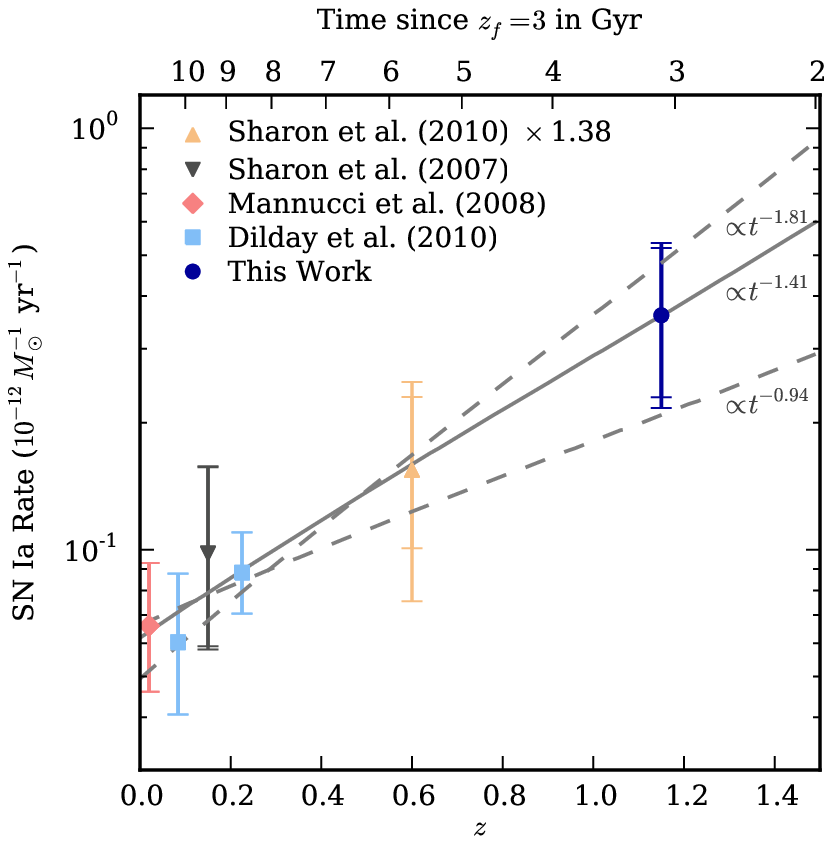}
\caption{Cluster rate measurements (all galaxy types) from this work
  and the literature. The rate of \citet{sharon10a} shown has been
  adjusted upward by 38\% from the reported rate (see text). The top
  axis shows the time elapsed since an assumed cluster formation
  redshift of $z_f = 3$. The \emph{solid grey line} represents the SN
  Ia rate for the best-fit power-law DTD: $\mathcal{R}_{\rm SN~Ia}(t)
  = \Psi(t)/m(t)$, where $\Psi(t) \propto t^s$. The \emph{dotted grey
    lines} show the range of $1\sigma$ error on $s$. 
\label{fig:clrates}}
\end{figure}

In Figure~\ref{fig:clrates} we compare our full cluster rate to the
lower-redshift rate measurements that have been normalized by stellar
mass, permitting a comparison across redshifts. Here we have made an
adjustment to the value reported by \citet{sharon10a}. Sharon et
al. used the mass-to-light ratio of Bell03 for the SDSS $g$ and $r$
bands, but did not apply a correction for evolution between $z \sim
0.6$ and $z = 0$.  Using the method described in \S\ref{lummass} we
find that a $-0.14$~dex offset should be applied to the mass to
account for evolution from $z=0.6$ to $z=0$ (Fig.~\ref{fig:mlratio},
right panel). We therefore adjust the reported rate of Sharon et
al. upward by 0.14~dex ($38\%$). The rate compilation of Maoz10
reflects this adjustment. Whereas the adjusted Sharon et al. rate
shows an indication that the cluster rate is increasing with redshift,
for the first time we find an increasing rate with high significance
($>2\sigma$).

We point out that the popular ``$A+B$'' model \citep{scannapieco05a}
is insufficient for describing the change in cluster rate with
redshift. In the $A+B$ model the SN rate is the sum of a term
proportional to the total stellar mass and a term proportional to the
recent star formation rate: $\mathcal{R}_{\rm SN~Ia} = AM_\ast + B
\dot{M}_\ast$. This simple model is convenient for predicting the SN
rate in environments with varying amounts of recent star formation as
it accounts for the increased SN~Ia rate at short delay times. (In
fact, we use this model in Meyers11 to derive limits on the expected
ratio of SNe~Ia to SNe~CC in early-type galaxies.) However, the model
lacks theoretical motivation and breaks down in other situations. For
example, \citet{greggio08a} note that it cannot adequately describe
the observed contribution from SNe with intermediate delay times
\citep[e.g.,][]{totani08a}.  This point is reinforced by the
observation of a changing cluster rate with redshift: In clusters, the
$A$ component is dominant at all redshifts observed. As $M_\ast$ is
not changing significantly with redshift, the rate would be expected
to remain constant under this model. Instead, we require a DTD model
wherein the rate decreases at large delay times (as it does in most
theoretically motivated models).

\vskip 0.4in
\subsection{The Cluster SN~Ia Delay Time Distribution} \label{conclusionsdtd}

The cluster rates constrain the SN~Ia delay time distribution,
$\Psi(t)$, over the range of delay times from a few Gyr to $\sim
10$~Gyr. To illustrate the cluster rate constraints, we parameterize
the DTD with a power law in time: $\Psi(t) \propto t^s$. A power law
is not only a convenient parameterization in the face of limited data,
but is a theoretically motivated function for the DD scenario, where
the late-time ($t \gtrsim 1$~Gyr) DTD shape is set by the distribution
of WD separation after the second CE phase and the merger timescale
due to gravitational radiation \citep{greggio05a}.

We make the approximation that all clusters formed in a single burst
of star formation at $z_f = 3$ and that the age of the stellar
population therefore corresponds to the elapsed time from $z_f$ to the
cluster redshift (Fig.~\ref{fig:clrates}, top axis). While
clearly a simplification, a single star-formation burst captures the
idea that the timescale over which star formation occured in cluster
early-type galaxies is short compared to the time since star formation
ceased.  The assumed burst redshift $z_f = 3$ is consistent with
measurements of cluster early-type galaxies showing that star
formation was mostly completed by this redshift
\citep[e.g.,][]{gobat08a}. Below, we show that the derived DTD is
relatively insensitive to the redshift assumed.

As noted in \S\ref{lummass}, the DTD is normalized by \emph{initial}
stellar mass, whereas the cluster rate measurements (including ours,
for consistency) have been normalized by \emph{current} stellar mass.
The DTD, $\Psi(t)$, is therefore related to the cluster rate by
$\Psi(t)=m(t)\mathcal{R}_{\rm SN~Ia}(t)$ where $m(t)$ is the fraction
of stellar mass remaining at time $t$ after the star formation
burst. The specific choice of $m(t)$ does not have a significant
impact on the derived DTD: regardless of the model or IMF assumed, the
stellar mass declines by only $\sim$10\% over the age range of
interest, $\sim 3$ to 11~Gyr. For consistency with Maoz10, we use the
remaining stellar mass fraction tabulated by BC03, $m_{\rm BC03}(t)$, but
corrected to $m(t) = 1 - (1 - m_{\rm BC03}(t))/0.7$ to effectively
convert from the Salpeter IMF used in BC03 to a ``diet'' Salpeter
IMF. This correction has only a very small effect on the result (see
below).

We find a best-fit value of
\begin{equation}
s = -1.41^{+0.47}_{-0.40},
\end{equation}
using the statistical$+$systematic error (added in quadrature)
reported for each rate measurement. In Figure~\ref{fig:clrates}, the
solid grey line shows the best-fit cluster rate for this value:
$\mathcal{R}_{\rm SN~Ia}(t)=\Psi(t)/m(t)$, where $\Psi(t) \propto
t^{-1.41}$. Note that the $\chi^2$ of the best-fit model is
surprisingly small: 0.40 for 4 degrees of freedom. The \emph{a priori}
probability of finding a $\chi^2$ smaller than 0.40 is less than
$2\%$. This is difficult to understand given that the measurement
errors are generally dominated by Poisson noise in the number of SNe
observed and are thus unlikely to be overestimated.

The best-fit value is consistent with measurements of the late-time DTD in
the field \citep{totani08a}. Most predictions for the SD scenario show
a steeper late-time DTD \citep{greggio05a,ruiter09a,mennekens10a} with
an effective value for $s$ ranging from $s \sim -1.6$
\citep{greggio05a} to $s < -3$ \citep{mennekens10a}, depending on the
details of the scenario and binary evolution. However, some groups
have found that the SD scenario could be consistent with a less-steep
DTD ($s \sim -1$) given the right combination of main sequence and red
giant secondaries \citep{hachisu08a}.  In the DD scenario, the
predicted shape of the DTD depends on the distribution of binary
separations after the common envelope phase of the WDs, a difficult
distribution to predict. However, a slope of $s = -1.4$ (and a range
of similar values) would not be surprising in the DD scenario.

\subsection{Additional DTD Systematic Uncertainties} \label{conclusionssys}

Variations in the assumed cluster star formation, initial mass
normalization and mass-to-light ratio evolution have a small affect on $s$
compared to the measurement error.

(1) \emph{Age of clusters' stellar populations:} Above, we assumed a
single burst of star formation at $z_f = 3$. Moving this single burst
to $z_f = 4$ results in $s = -1.55$. A more recent burst, $z_f = 2.5$,
results in $s = -1.30$.  Maoz10 give a treatment of variations from
the single-burst approximation, also finding that the affect on $s$ is
small.

Our rate measurements in red and early-type galaxies provide a good
consistency check that recent star formation does not significantly
contribute to the SN~Ia rate: if it did, we would observe a higher
rate in the full cluster than in these subsamples. Surprisingly, we
observe the opposite trend (although the significance is low). The
red-sequence early-type subsample includes 53\% of the stellar mass of
the full cluster sample, and 6 SNe~Ia. The remaining 47\% of the full
cluster sample (which includes bluer galaxies and late-type
red-sequence galaxies) accounts for only $2 \pm 1$ SNe~Ia. At low
redshift, \citet{mannucci08a} found a similar trend between E/S0
galaxies and S0a/b galaxies within $0.5$~Mpc of cluster centers,
though also at $<1\sigma$ significance.

(2) \emph{Remaining stellar mass:} Whereas the DTD is normalized by
initial stellar mass and cluster rate measurements have been
normalized by current stellar mass, we have assumed a remaining
stellar mass fraction $m(t)$ to convert from current to initial
stellar mass. Although different models and IMFs can yield sigificantly
different $m(t)$, we are only concerned here with the change in $m(t)$
between $\sim 3$~Gyr and at $\sim 11$~Gyr. (The absolute value of
$m(t)$ affects only the normalization of $\Psi(t)$, with which we are
not concerned.) Fortunately, the evolution in $m(t)$ in this age range
is small and consistent between models, and so the effect on $s$ is
small. For example, using $m_{\rm BC03}(t)$ (assuming a Salpeter IMF)
rather than correcting to a diet Salpeter IMF (as we have done) only
changes the best-fit value from $s=-1.41$ to $s=-1.38$.

If in \S\ref{lummass} we had used a $M/L$ ratio directly normalized by
initial mass, rather than normalizing by current mass and later
converting to initial mass, the results would be very
similar. (We have not done this for consistency with other
rate measurements.) In the {\sc p\'egase2} models in
Figure~\ref{fig:mlratio} (left panel) evaluted at $z=1.2$, the
ratio of current to formed stellar mass varies slightly across the
models, but is fully contained in the range $0.66 \pm 0.03$. The same
models evaluted at $z=0$ have a ratio of $0.59 \pm 0.03$. This is
consistent with the $\sim 10\%$ evolution in $m(t)$ over this range as
tabluated by BC03.

(3) \emph{$M/L$ ratio evolution:} While the overall normalization of
the $M/L$ ratio will only affect the normalization of $\Psi(t)$ and
not $s$, the evolution of the $M/L$ ratio will affect $s$. In
\S\ref{lummass} we assigned a liberal 20\% systematic uncertainty to
the evolution of the $M/L$ ratio over the redshift range of
interest. To estimate the effect of this systematic uncertainty, we
adjust our rate measurement by 20\% and that of \citet{sharon10a} by
10\% and refit $s$. The resulting change in $s$ for positive and
negative shifts is $-0.15$ and $+0.18$ respectively, less than half of
the nominal error in $s$.

\vskip 0.5in
\subsection{Conclusions}

In this paper, we have made a measurement of the high-redshift cluster
SN~Ia rate. Thanks to an unusually complete dataset (particularly for
a rate study) the measurement is quite robust, with statistical and
systematic uncertainties on par with or better than measurement
uncertainties at low redshift. We highlight several important and/or
unique aspects of the measurement:
\begin{itemize}
\item The SN classification approach takes advantage
  of all relevant information. Thanks to the ``rolling search''
  strategy of the survey and the nearly complete spectroscopic
  follow-up, most candidates have a full light curve and a host galaxy
  redshift, greatly reducing classification uncertainty.
\item The position-dependent control time allows one to calculate a
  supernova rate given an arbitrary observing pattern and luminosity
  distribution.
\item The control time calculation includes a full distribution of SN
  properties and the systematic uncertainty associated with the assumed 
  distribution is carefully quantified. Thanks to the
  depth of the observations, the detection efficiency approaches
  100\% during the period of the survey for most of the clusters,
  meaning that the systematic uncertainty is low.
\item Statistical uncertainties associated with the cluster
  luminosities, including both statistical variations and cosmic
  variance, are included in the total uncertainty.  Also, light in the
  outskirts of each galaxy (outside the {\sc SExtractor} MAG\_AUTO
  aperture) is accounted for.  This is a significant component of the
  total cluster luminosity.
\item Cluster SN~Ia rate measurements are normalized consistently across
  redshifts using a redshift-dependent mass-to-light versus color
  relation.
\end{itemize}

For the first time our result shows at the $>2\sigma$ level that the
cluster SN~Ia rate is increasing with redshift.  Simply by comparing
the low- and high-redshift cluster rate measurements, the shape of the
late-time SN~Ia delay time distribution can be constrained. The
power of the measurement for this purpose comes both from
the high redshift and relatively low statistical and systematic
uncertainties in the measurement. While we cannot conclusively rule
out either the single degenerate or double degenerate class of
progenitors via the delay time distribution, the binary
evolution that could lead to each model are constrained. The
DD scenario is consistent with the measurement under a wide range of
plausible binary evolution parameters, while there is a stronger
constraint on binary scenarios that could lead to an SD
scenario. Finally, this measurement is unique in constraining the delay
time distribution at delay times of a few Gyr. In future studies, it
can be used in combination with other cluster rates and other delay
time distribution measurements (e.g., Maoz10) to place even tighter
constraints on models for binary evolution and SN~Ia progenitor
scenarios.

\acknowledgements
We thank Eric Bell and Dan Maoz for helpful discussion. T.~M. is
financially supported by the Japan Society for the Promotion of
Science (JSPS) through the JSPS Research Fellowship.  C.~L. is
financially supported by the Australian Research Council (ARC) through
the ARC Future Fellowship program.  Financial support for this work
was provided by NASA through program GO-10496 from the Space Telescope
Science Institute, which is operated by AURA, Inc., under NASA
contract NAS 5-26555.  This work was also supported in part by the
Director, Office of Science, Office of High Energy and Nuclear
Physics, of the U.S. Department of Energy under Contract
No. AC02-05CH11231, as well as a JSPS core-to-core program
``International Research Network for Dark Energy'' and by a JSPS
research grant (20040003).  The authors wish to recognize and
acknowledge the very significant cultural role and reverence that the
summit of Mauna Kea has always had within the indigenous Hawaiian
community.  We are most fortunate to have the opportunity to conduct
observations from this mountain.  Finally, this work would not have
been possible without the dedicated efforts of the daytime and
nighttime support staff at the Cerro Paranal Observatory.
{\it Facilities:} \facility{HST (ACS)}, \facility{Subaru
     (FOCAS)}, \facility{Keck:I (LRIS)}, \facility{Keck:II
     (DEIMOS)}, \facility{VLT:Antu (FORS2)}

\end{document}